\documentclass[sigconf]{acmart}

\usepackage{subcaption}
\usepackage{tcolorbox}
\usepackage{lipsum}
\usepackage{multirow}
\usepackage{colortbl} 
\usepackage{xcolor}
\tcbuselibrary{skins, breakable, theorems}
\usepackage{tabularx}
\usepackage{booktabs}
\usepackage{threeparttable}
\usepackage{caption}
\usepackage{array}
\AtBeginDocument{%
  }

\setcopyright{acmlicensed}
\copyrightyear{2018}
\acmYear{2018}
\acmDOI{XXXXXXX.XXXXXXX}

\begin{document}

\title{Exploring a Gamified Personality Assessment Method through Interaction with LLM Agents Embodying Different Personalities}

\author{Baiqiao Zhang}
\affiliation{%
  \institution{The Hong Kong University of Science and Technology}
  \state{Hong Kong SAR}
  \country{China}
}
\email{baiqiao.zhang@connect.ust.hk}

\author{Xiangxian Li}
\affiliation{%
  \institution{Shandong University}
  \state{Shandong}
  \country{China}
}
\email{xiangxianli@sdu.edu.cn}

\author{Chao Zhou}
\affiliation{%
  \institution{Institute of Software Chinese Academy of Sciences}
  \city{Beijing}
  \country{China}
}
\email{zhouchao@iscas.ac.cn}

\author{Xinyu Gai}
\affiliation{%
  \institution{The Hong Kong University of Science and Technology (Guangzhou)}
  \state{Guangzhou}
  \country{China}
}
\email{xge236@connect.hkust-gz.edu.cn}

\author{Juan Liu}
\affiliation{%
  \institution{Shandong University}
  \state{Shandong}
  \country{China}
}
\email{zzzliujuan@sdu.edu.cn}

\author{Xue Yang}
\affiliation{%
  \institution{Tsinghua University}
  \state{Beijing}
  \country{China}
}
\email{yang_shirley@tsinghua.edu.cn}

\author{Nianlong Li}
\affiliation{%
  \institution{Institute of Software Chinese Academy of Sciences}
  \city{Beijing}
  \country{China}
}
\email{linianlong@iscas.ac.cn}

\author{Shuai Ma}
\affiliation{%
  \institution{Institute of Software Chinese Academy of Sciences}
  \city{Beijing}
  \country{China}
}
\email{mashuai@iscas.ac.cn}

\author{Xiaojuan Ma}
\affiliation{%
  \institution{The Hong Kong University of Science and Technology}
  \state{Hong Kong SAR}
  \country{China}
}
\email{mxj@cse.ust.hk}

\author{Yong-jin Liu}
\authornote{Corresponding Authors}
\affiliation{%
  \institution{Tsinghua University}
  \state{Beijing}
  \country{China}
}
\email{liuyongjin@tsinghua.edu.cn}

\author{Yulong Bian}
\authornotemark[1]
\affiliation{%
  \institution{Shandong University}
  \state{Shandong}
  \country{China}
}
\email{bianyulong@sdu.edu.cn}

\renewcommand{\shortauthors}{Zhang et al.}

\begin{abstract}
The low-intrusion and automated personality assessment is receiving increasing attention in psychology and human-computer interaction fields. This study explores an interactive approach for personality assessment, focusing on the multiplicity of personality representation. We propose a framework of Gamified Personality Assessment through Multi-Personality Representations (Multi-PR GPA). The framework leverages Large Language Models to empower virtual agents with different personalities. These agents elicit multifaceted human personality representations through engaging in interactive games. Drawing upon the multi-type textual data generated throughout the interaction, it achieves personality assessments with interpretable insights. Grounded in the classic Big Five personality theory, we developed a prototype system and conducted a user study to evaluate the efficacy of Multi-PR GPA. The results affirm the effectiveness of our approach in personality assessment and demonstrate its superior performance when considering the multiplicity of personality representation. Error structure analysis further revealed systematic assessment biases in LLMs, which multi-context aggregation partially mitigated.

\end{abstract}

\begin{teaserfigure}
  \includegraphics[width=\textwidth]{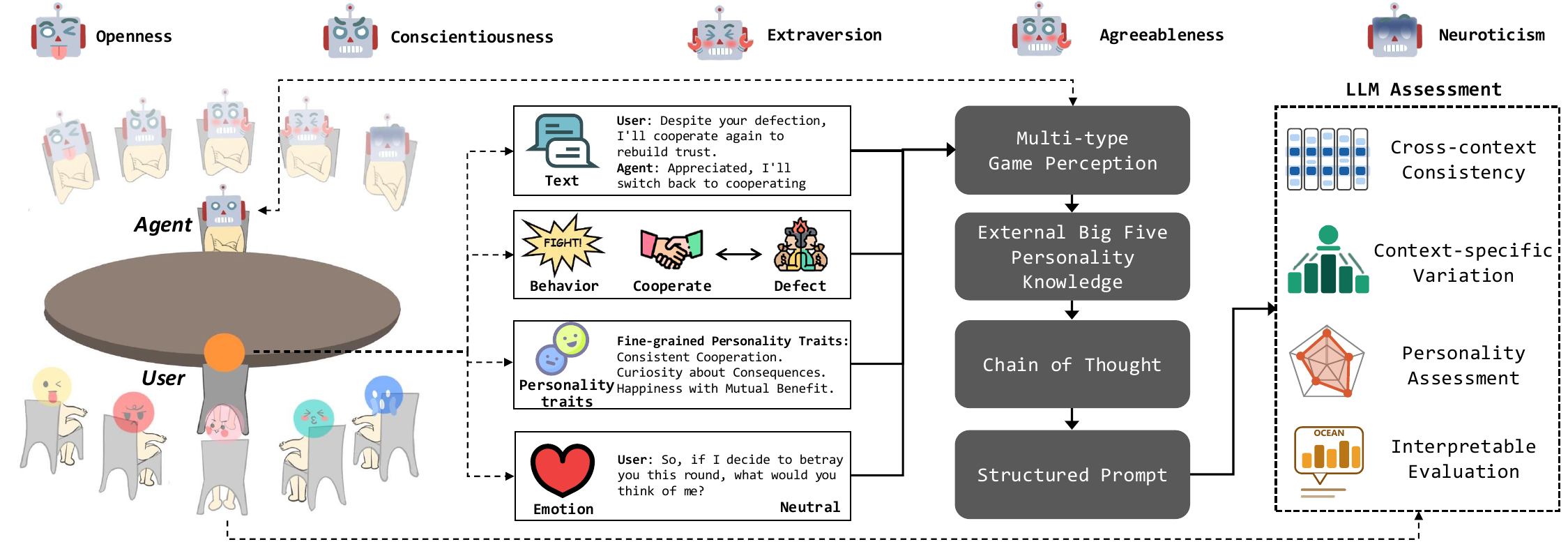}
  \caption{A gamified personality assessment framework through interacting with agents embodying different personalities. The top icons represent agents' personalities; the bottom icons represent the user's multifaceted personality representations.}
  \label{fig:teaser}
\end{teaserfigure}
\received{20 February 2007}
\received[revised]{12 March 2009}
\received[accepted]{5 June 2009}

\maketitle

\section{Introduction}
\label{Intro}

Understanding personality --- a dynamic organization of psychophysical systems that creates characteristic and consistent patterns of human behavior, cognition, and feelings \cite{allport1936trait} --- is crucial for user-centered interactive system designs. Personality influences how users interact with computing systems, from interface preferences and learning styles \cite{zhu2023influence, subaramaniam2022users, stachl20192} to long-term engagement patterns \cite{sorman2024influence, akhtar2015engageable}. However, accurately assessing personality remains challenging. Self-report questionnaires (e.g., BFI-44 \cite{john1999big}) are the most established instruments, yet their results may be influenced by social desirability bias \cite{kreitchmann2019controlling}. Projective tests and situational assessments attempt to mitigate this through indirect observation, but rely heavily on administrators' subjective interpretation and scarce expert resources \cite{mussel2016situational, crisp2014global}. More recently, machine learning approaches have been applied to infer personality from static text such as social media posts \cite{peters2024large2, mairesse2007using}. While promising, these methods analyze pre-existing data in a single context and thus provide only a limited window into an individual's personality.


This is a shortcoming due to the \textit{multiplicity} of personality: individuals naturally present different facets depending on the situation and the people they interact with \cite{apa_multipleSelves, mischel1973toward, snyder1985personality} (see Fig.~\ref{fig:teaser}). The Cognitive-Affective Personality System (CAPS) identifies the interaction partner's characteristics as a core psychologically active feature that activates different \textit{if-then} behavioral signatures \cite{mischel1995cognitive}. An individual may respond cautiously to a confrontational partner, warmly to a cooperative one, and competitively to a risk-seeking one. In this paper, we define \textit{interaction context} as the social situation characterized by the interaction partner's personality, such that interacting with partners of different personalities constitutes different contexts within the same environment. Observing someone within only one interaction context risks capturing a partial, potentially misleading picture of their personality. Comprehensive assessment therefore requires observation across multiple interaction contexts through sufficient engagement to allow personality-relevant patterns to emerge. Yet existing approaches, including gamified assessments \cite{harman2024advances, weidner2019playing, wu2022individual} and dialogue-based systems \cite{lee2024chatfive, yang-etal-2024-psychogat}, have not incorporated such multi-context observation.

Operationalizing such an observation has been impractical with conventional methods: recruiting multiple human interaction partners with controlled personality profiles is costly and difficult to standardize. Recent advances in large language models (LLMs) offer a viable path. LLM-powered agents can be endowed with distinct, controllable personality traits \cite{jiang2024evaluating, serapio2023personality, huang2023revisiting}. According to the Computers Are Social Actors (CASA) paradigm \cite{nass1994computers}, people unconsciously apply the same social heuristics to computers as to humans, meaning that users interacting with personality-driven agents can naturally reveal their behavioral tendencies. Deploying multiple such agents with different personalities thus creates the diverse social contexts needed for multi-situational personality observation.

Building on this insight, we propose a framework of Gamified Personality Assessment through Multi-Personality-Representations (Multi-PR GPA). The framework uses multiple LLM agents, each embodying a dominant Big Five personality trait, as interaction partners in a Prisoner's Dilemma-based trust game. By comparing user response patterns across these agents, it captures both the \textbf{consistency} and \textbf{variation} in each user's personality expression across contexts. We developed a prototype system and conducted a user study with 42 participants, each engaging in multi-round dialogue and strategic decision-making with five personality-driven agents in a randomized order. We evaluated the framework through technical evaluation using six LLMs as personality evaluators, prediction error structure decomposition, and qualitative case analysis. 

\textbf{Our contributions are as follows:}

\begin{itemize}
    \item We propose Multi-PR GPA, a novel framework that incorporates the multiplicity of personality representation into automated personality assessment, and validate it through a user study with 42 participants across six LLMs. Multi-context assessment consistently outperformed all single-context baselines, and participants reported good engagement throughout the interaction process.
    \item Through error structure decomposition and qualitative analysis, we reveal that LLMs exhibit systematic directional biases when scoring personality, and multi-context aggregation partially mitigates these biases by reducing errors in both directions simultaneously.
    \item We derive design implications for interactive personality assessment systems, including the prioritization of context diversity over depth in a single context, the need for dimension-specific calibration of LLMs, and the integration of diverse interaction tasks to improve coverage of underrepresented dimensions.
\end{itemize}




\section{Related Works}
\label{sec:related}
\subsection{Personality and Its Multiplicity}

The Big Five model proposes five trait dimensions --- Openness, Conscientiousness, Extraversion, Agreeableness, and Neuroticism --- and is the most widely adopted framework for personality structure \cite{goldberg2013alternative, costa1999five}. Validated instruments such as BFI-44 \cite{john1999bigfive} have been developed to measure these dimensions. Empirical research shows that Big Five traits are associated with distinct behavioral tendencies \cite{john1999big}, language use patterns \cite{pennebaker1999linguistic}, and emotional responses \cite{revelle2009personality}. These associations between personality and observable signals in behavior, language, and emotion provide a basis for inferring personality from interaction data.

An individual's behavior in social contexts is shaped by both dispositional characteristics and situational factors \cite{snyder1985personality}. The Cognitive-Affective Personality System (CAPS) \cite{mischel1995cognitive} formalizes this through \textit{if-then} behavioral signatures: stable patterns in how individuals respond differently to different situational features. This means personality does not manifest as fixed behavioral output but as context-sensitive patterns --- the \textit{multiplicity} of personality representation \cite{apa_multipleSelves, mischel1973toward}. CAPS implies that a single context activates only one conditional expression of the personality system \cite{mischel1995cognitive}. Identifying stable dispositional traits therefore requires observing how an individual's responses vary systematically across contexts \cite{diener1984person}. Guided by this theoretical foundation, our framework exposes users to multiple distinct interaction contexts to capture the cross-context behavioral patterns that reveal underlying personality traits.

\subsection{Personality Assessment}
Methods for personality assessment have evolved from static self-report measures toward increasingly data-rich approaches.

\textbf{Self-report and indirect assessment.} Self-report questionnaires such as BFI-44 \cite{john1999big} remain the standard instrument for personality measurement, though they may be influenced by social desirability bias \cite{kreitchmann2019controlling}. Projective tests (e.g., Rorschach \cite{rorschach1942psychodiagnostics}) and situational judgment tests use indirect measurement to reduce this influence, but depend on administrators' subjective interpretation \cite{mussel2016situational} and scarce expert resources \cite{crisp2014global}.

\textbf{Machine learning on static data.} Automated approaches have applied machine learning to various personality-relevant data sources, including voice signals \cite{gilpin2018perception}, online and offline behavioral data \cite{kim2018detecting}, and eye-tracking data \cite{berkovsky2019detecting}. More recently, LLMs have been used to infer personality from user-generated text such as social media posts \cite{peters2024large2, wright2026assessing,zhang2024can,peters2024large}. Integrating psychological knowledge has further improved prediction accuracy, for example by combining prompts with questionnaire items \cite{yang2023psycot} or incorporating emotion regulation theory via RAG \cite{li2024eerpd}. However, these approaches predominantly analyze static, pre-existing data from a single context.

\textbf{Gamified and interactive assessment.} Game-Based Assessment (GBA) introduces interactive elements to enhance participant engagement and reduce resistance to testing \cite{gomez2022systematic, mccord2019game}. Studies have shown that in-game behavior correlates with Big Five dimensions \cite{weidner2019playing, wu2022individual}, and that serious games can achieve validity comparable to standard questionnaires \cite{ramos2024serious}. More recent work explores LLM-driven interaction for assessment. PsychoGAT \cite{yang-etal-2024-psychogat} wraps traditional questionnaire items in narrative scenarios, but users still select from predefined options, preserving the underlying questionnaire paradigm. ChatFive \cite{lee2024chatfive} uses guided LLM dialogue to assess Big Five traits. However, all these systems operate within a single interaction context with a uniform agent persona.

\begin{figure*}[!t]
    \centering
	\includegraphics[width=2\columnwidth]{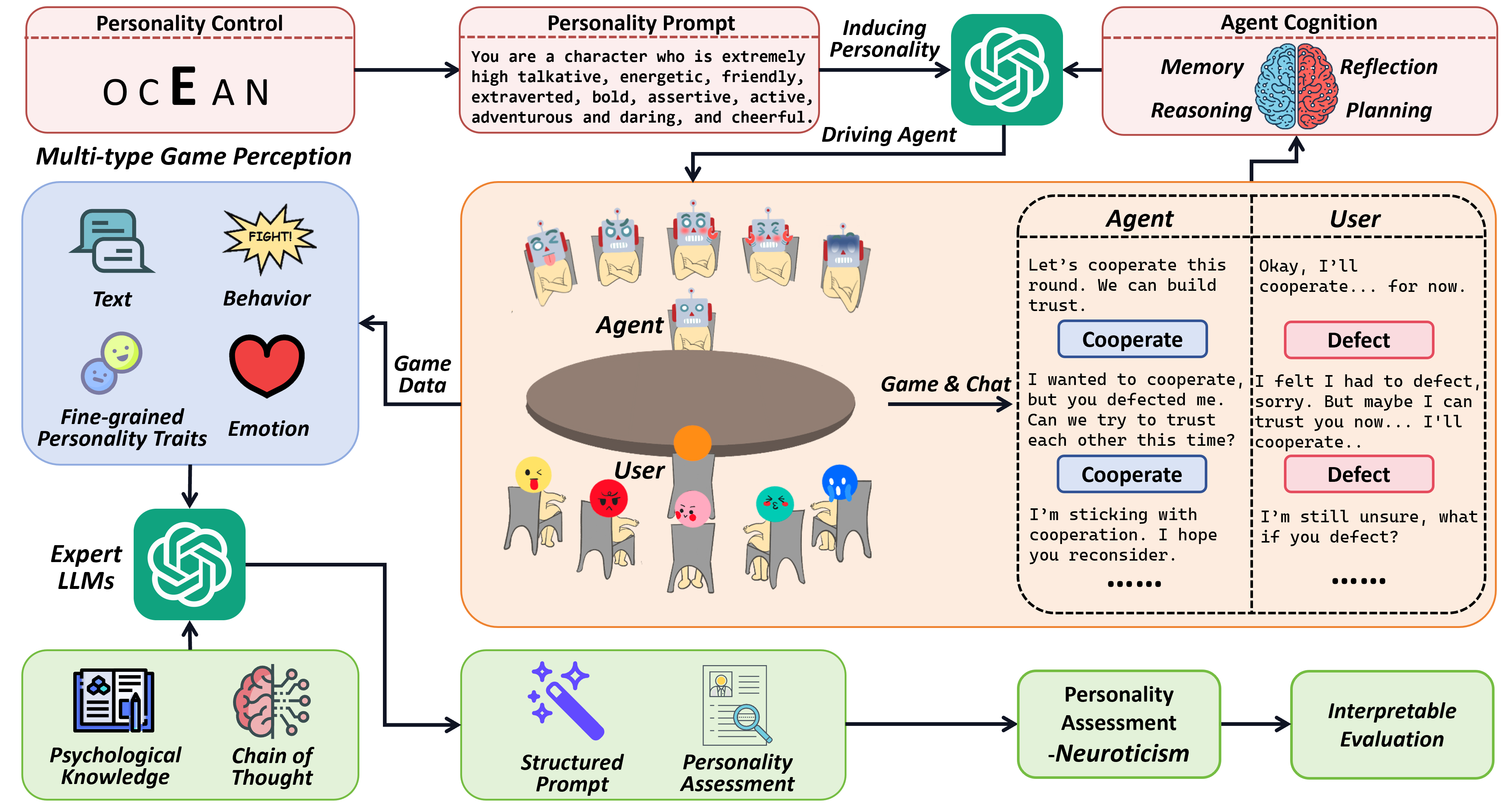}
	\caption{The framework includes Gamified Interaction (orange block, section \ref{GI}), LLM Agent with Controlled Personality (pink block, section \ref{LAMP}), Multi-type Game Data Perception (blue block, section \ref{MTTP}), and Personality Assessment (green block, section \ref{PA}). This framework assesses personality based on the multi-personality representations during interactions.}
	\label{fig:fig2}

\end{figure*}

\textbf{Remaining gaps.} Two gaps persist across existing approaches. First, no method for human personality assessment systematically varies the interaction context to capture the multiplicity of personality expression --- assessments operate within a single situation with a single type of interaction partner. Second, interactive approaches have not fully utilized the multi-type data that rich interaction naturally generates, typically relying on a single data modality. Our framework aims to address both gaps.

\subsection{LLM Agents with Controllable Personality}

Creating agents with specific, controllable personality traits is a prerequisite for our multi-context framework. LLMs, trained on large-scale corpora containing language from individuals with diverse personalities, can be guided to exhibit targeted traits through prompt engineering or fine-tuning \cite{jiang2024evaluating}.

Prompt-based methods are widely used: Jiang et al. \cite{jiang2024evaluating} induced Big Five traits in ChatGPT through prompt design; Huang et al. \cite{huang2023revisiting} showed that \texttt{gpt-3.5-turbo} exhibits varied traits under specific prompt adjustments; Serapio-Garcia et al. \cite{serapio2025psychometric} used Likert-scale language qualifiers for finer control over trait levels. Fine-tuning offers an alternative: Liu et al. \cite{liu2024dynamic} combined LoRA with hypernetworks to generate personality-specific adapter weights dynamically.

The previous works confirm that LLMs can serve as believable agents with diverse, controllable personalities. We aim to utilize this capability to create distinct interaction partners, each showing a dominant Big Five trait, thereby providing the situational diversity needed to elicit the multiplicity of users' personality expressions.

\section{Framework Design and Implementation}
\label{sec:system}

Guided by the theoretical foundation that personality assessment benefits from multi-situational observation (Section~\ref{sec:related}), we propose a framework of Gamified Personality Assessment through Multi-Personality-Representations (Multi-PR GPA), shown in Fig.~\ref{fig:fig2}. The framework contains four components: Gamified Interaction, LLM Agents with Controlled Personality, Multi-Type Game Perception, and Personality Assessment. We take the Big Five model as the reference given its extensive cross-cultural validation \cite{matthews2003personality, gurven2013universal}. Below we describe each component along with its implementation.

\subsection{Gamified Interaction}
\label{GI}
 
The gamified interaction component serves two purposes: enabling natural expression of thoughts and emotions, and stimulating social behaviors and psychological reasoning that externalize personality traits through language and decision-making.
 
\begin{figure}[t]
    \centering
        \begin{subfigure}[b]{0.23\textwidth}
        \centering
        \includegraphics[width=\textwidth]{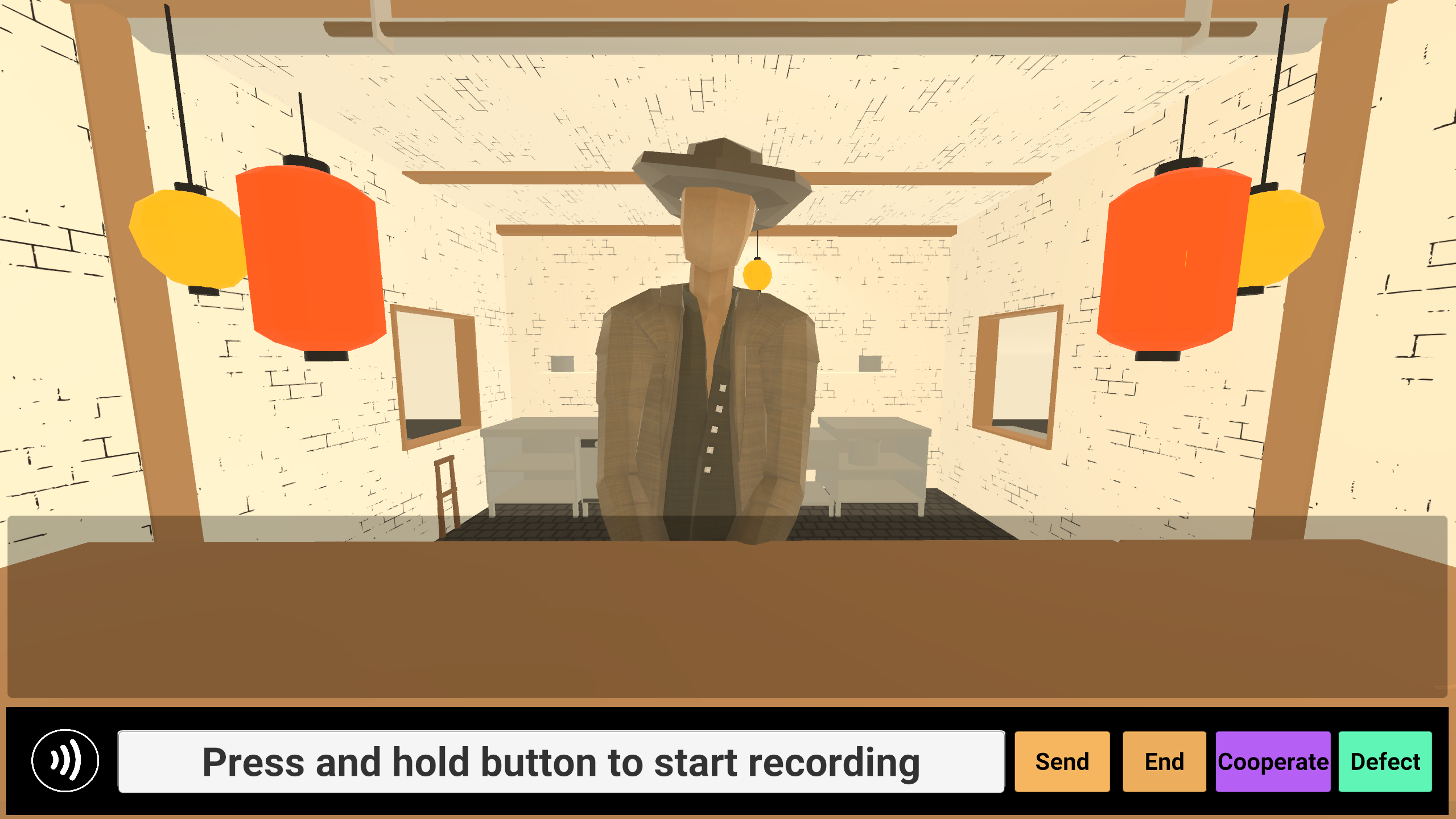}
        \label{fig:4a}
    \end{subfigure}
    \hfill
    \begin{subfigure}[b]{0.23\textwidth}
        \centering
        \includegraphics[width=\textwidth]{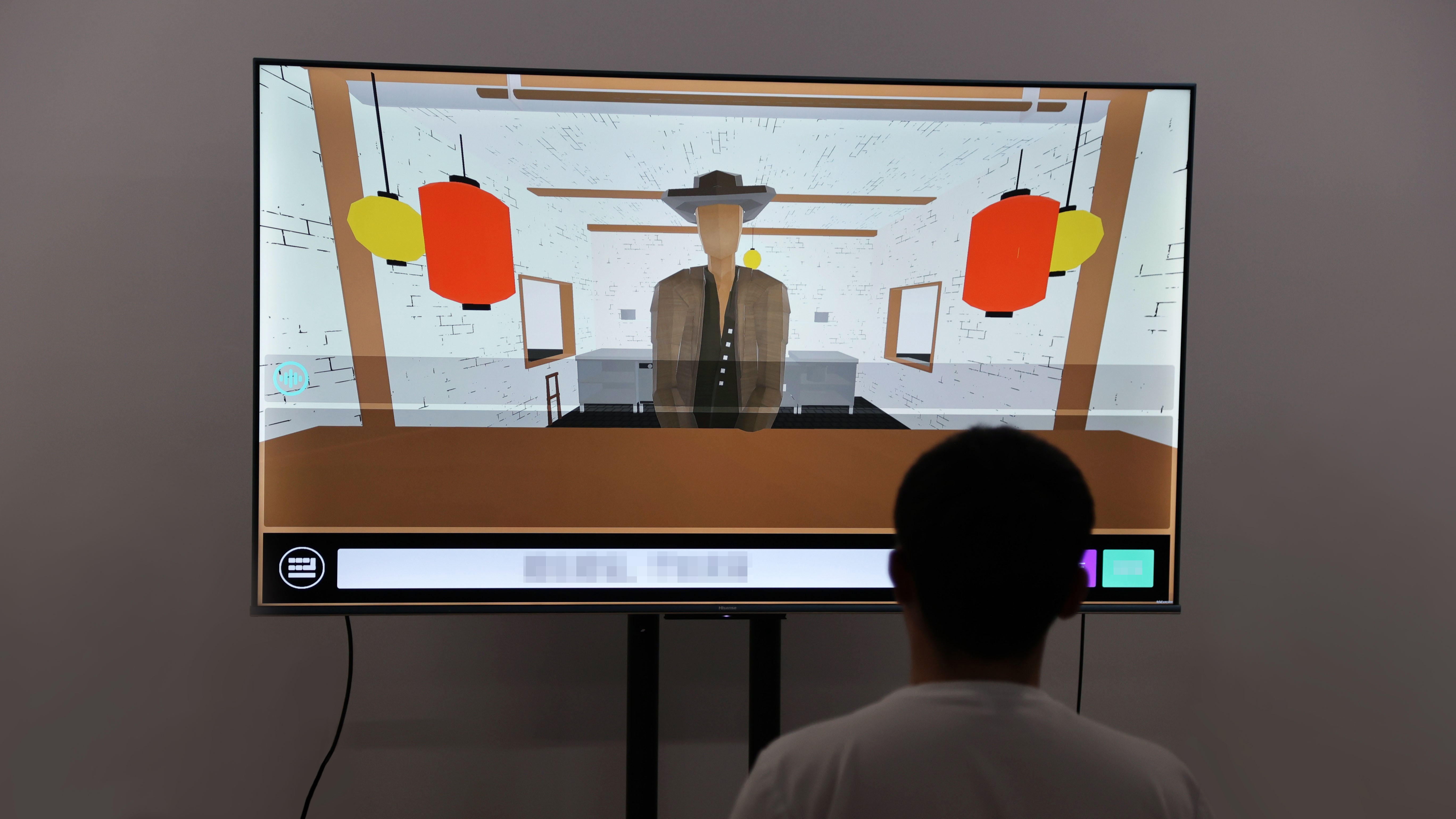}
        \label{fig:3b}
    \end{subfigure}
    \caption{The prototype system used in the experiment.}
    \label{fig:3}
\end{figure}

Personality assessment requires a game environment where strategic social interaction is central, so that users' decisions reflect their authentic tendencies rather than task-specific skills. The Prisoner's Dilemma satisfies this requirement: its cooperate-or-defect mechanic inherently involves trust calibration, risk tolerance, and fairness reasoning --- behaviors closely tied to personality traits --- and has a solid empirical foundation in behavioral psychology \cite{axelrod1981evolution, fehr2002altruistic}. We adapted the variant ``The Evolution of Trust'' \cite{case2015evolution} into a multi-round game where each round comprises a \textit{dialogue phase} (free-form conversation with the agent via voice or text) followed by a \textit{decision phase} (independent cooperate/defect choice). To encourage authentic self-expression rather than strategic optimization, we designed a narrative storyline that de-emphasizes win/loss outcomes (see Appendix~\ref{sec:story}) and hid game scores and round counts from the interface \cite{jia2016personality}. The system is shown in Fig. \ref{fig:3}.

\subsection{LLM Agents with Controlled Personality}
\label{LAMP}
 
To create the diverse interaction contexts needed for multi-situational observation, we deploy five LLM agents, each with a dominant Big Five trait induced through personality prompts (shown in the pink block of Fig.~\ref{fig:fig2}). According to the CASA paradigm \cite{nass1994computers}, users apply social heuristics to computer agents as they would to humans, so agents exhibiting distinct personalities can serve as varied social interaction partners.
 
However, personality prompts alone are insufficient for multi-round interaction: a static prompt cannot adapt to unfolding game dynamics while maintaining personality consistency. We therefore equip each agent with a cognition module that enables self-evolving behavior across rounds. The module comprises four sub-modules that form an iterative loop: \textbf{Memory} accumulates game outcomes and interaction patterns; \textbf{Reflection} evaluates these experiences through the lens of the agent's assigned personality; \textbf{Reasoning} integrates memory and reflection to make personality-consistent decisions, aided by Chain-of-Thought (CoT) prompting \cite{wei2022chain} and Theory of Mind (ToM) \cite{strachan2024testing}; and \textbf{Planning} generates strategies for subsequent rounds. This loop allows each agent to dynamically adapt to the user's behavior while maintaining coherent personality expression --- for instance, an agreeable agent that has been repeatedly defected against will reflect on the broken trust and adjust its cooperative strategy, whereas a neurotic agent facing the same situation will escalate its anxiety-driven responses. Detailed prompts are provided in Appendices~\ref{memory}--\ref{reasoning}.

\subsection{Multi-Type Game Perception}
\label{MTTP}
 
To capture personality signals at multiple levels of granularity, we extract four types of textual data from each interaction with agent $a_k$ ($k \in \{O, C, E, A, N\}$):
 
\begin{equation}
D_k = \{T_k, B_k, E_k, P_k\}
\end{equation}
 
\noindent where $T_k$ denotes the natural dialogue text and $B_k$ the user's cooperate/defect decision sequence. These two constitute foundational data collected during gameplay. $E_k$ and $P_k$ are higher-level abstractions extracted post-hoc to enrich the personality signal:

\noindent $E_k$: emotion labels classified from each user utterance via zero-shot inference with \texttt{GPT-5.4}. We validated this approach on the SMP2020-EWECT benchmark~\cite{smp2020ewect} (see Appendix~\ref{sec:emotion_eval}).

\noindent $P_k$: fine-grained personality traits inferred from each round by an LLM configured as a Big Five expert. The purpose of $P_k$ is to extract salient personality-relevant details from fragmented interactions, mitigating attention dilution that occurs when LLMs process extended multi-round contexts directly.

\noindent To ensure fairness, only foundational data ($T_k$, $B_k$) is provided to agents during the game; higher-level data ($E_k$, $P_k$) is reserved exclusively for personality assessment. Extraction prompts are detailed in Appendices~\ref{emotion} and~\ref{Fine-grained}.

\subsection{Personality Assessment}
\label{PA}

The complete observation across all five agents forms:

\begin{equation}
\mathcal{D} = \{D_O, D_C, D_E, D_A, D_N\}
\end{equation}

\noindent The personality assessment is then formulated as:

\begin{equation}
\hat{\mathbf{y}} = f_{\text{LLM}}(\mathcal{D}, \mathcal{K}, \mathcal{P}_{\text{CoT}})
\end{equation}

\noindent where $\hat{\mathbf{y}} \in \mathbb{R}^5$ is the predicted Big Five score vector, $\mathcal{K}$ denotes expert psychological knowledge describing the behavioral indicators associated with each Big Five dimension (Appendix~\ref{app:expert_knowledge}), and $\mathcal{P}_{\text{CoT}}$ denotes the structured Chain-of-Thought instructions. A neutral LLM serves as $f_{\text{LLM}}$ to minimize evaluator bias.

The CoT instructions $\mathcal{P}_{\text{CoT}}$ guide the LLM through a three-step reasoning process: (1) \textit{cross-context consistency analysis} --- identifying behavioral patterns that remain stable across all five agents, which likely reflect core dispositional traits; (2) \textit{context-specific variation analysis} --- examining how the user's behavior changes in response to different agent personalities (e.g., becoming more cooperative with agreeable agents but more defensive with neurotic agents); and (3) \textit{synthesis} --- integrating consistent patterns and meaningful variations to produce a holistic score for each dimension, accompanied by an interpretable rationale linking the score to observed behavioral evidence.

This formulation supports our experimental comparisons. Using a single $D_k$ in place of $\mathcal{D}$ yields the single-context baselines, and removing $E_k$ or $P_k$ from each $D_k$ corresponds to the ablation conditions. The complete assessment prompt is provided in Appendix~\ref{sec:personality_assess}.

\subsection{Implementation Details}
All agent cognition tasks during gameplay were performed using \texttt{gpt-4o-0806} with temperature set to 0. Post-hoc extraction of emotion labels and fine-grained personality traits was performed using \texttt{GPT-5.4}. The personality assessment module was evaluated using six different LLMs to test the framework's generalizability (detailed in Section~\ref{sec:settings}). Further details on the game engine, speech processing, and interface design are provided in Appendix~\ref{app:impl_details}.

\section{User Study}

To validate the proposed Multi-PR GPA framework, we conducted a user study with 42 participants. Our evaluation consists of three parts: (1) \textit{User Experience Survey}, which examines user engagement and naturalness during gamified interaction; (2) \textit{Technical Evaluation}, which compares multi-context against single-context assessment across multiple LLMs and two independent ground truths (self-report and expert ratings), with ablation study and data-volume-controlled comparisons to isolate the sources of improvement; and (3) \textit{Error Structure Analysis}, which decomposes prediction errors into overestimation and underestimation components to understand the sources of assessment bias; and (4) \textit{Qualitative Analysis}, which illustrates how multi-context observation improves assessment at the participant level.

\begin{figure*}[ht]
    \centering
	\includegraphics[width=2\columnwidth]{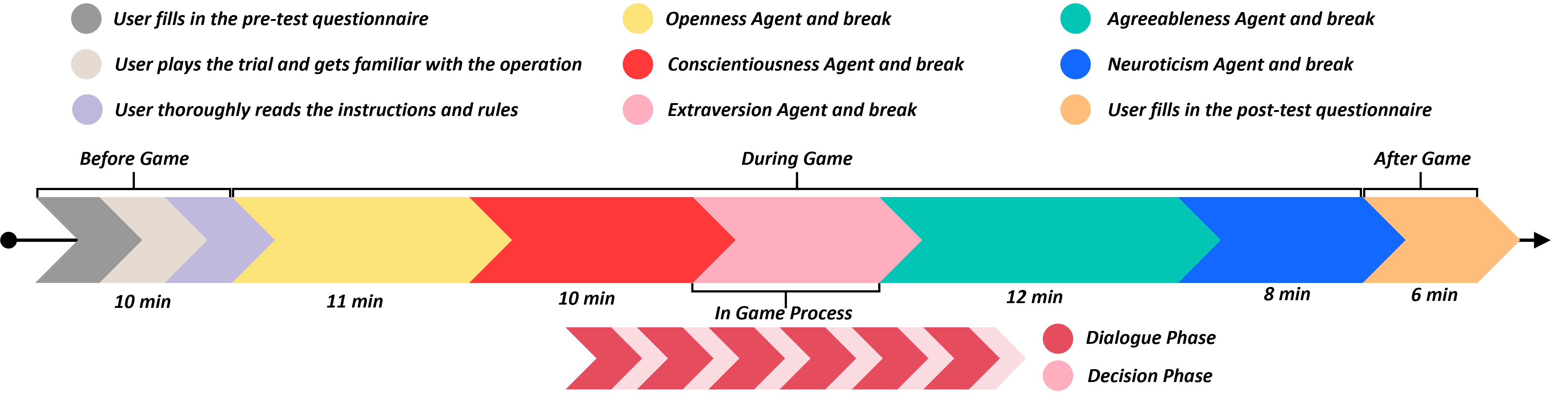}
	\caption{Overview of the experimental procedure.}
	\label{fig:fig4}
\end{figure*}

\subsection{Participants}
\label{sec:participants}

We first conducted a power analysis to determine the required sample size using G*Power \cite{faul2009statistical}. With an effect size $d_z = 0.5$ (indicating a medium effect), significance threshold $\alpha = 0.05$, and statistical power $1-\beta = 0.8$, the results indicated a total sample size of 27 was needed for the two conditions. We recruited 42 participants (21 males, 21 females; $M = 22.07$, $SD = 2.32$ years) from a local university. Post-hoc power analysis revealed an achieved statistical power of $1-\beta = 0.938$. Individuals who had consumed alcohol, experienced severe fatigue, taken medication, or been ill immediately before the experiment were excluded. This study adhered to the Declaration of Helsinki and was approved by the Human Research Ethics Committee. All participants gave written informed consent after being told about the general procedures. To avoid bias, this study employed a post-study disclosure design. Participants were blinded to the specific purpose (assessing personality traits) during the experiment and were debriefed after the experiment. During debriefing, participants were fully informed, compensated \$10, and given the option to confirm or withdraw data usage consent. All ultimately agreed to the use of their data for research.


\subsection{Task and Procedure}

We conducted the experiment using the implemented prototype system described in Section~\ref{sec:system}. To ensure manageable study duration and prevent participant fatigue, we set the number of interaction rounds with each agent to six based on small-scale user testing in the development phase, where interactions lasted around ten minutes, keeping users engaged without boredom. As illustrated in Fig. \ref{fig:fig4}, the experimental procedure consisted of three main phases:

\textbf{Before Game Phase:} Participants first completed the BFI-44 questionnaire, then engaged in a practice session to familiarize themselves with the system operation. They were subsequently instructed to carefully read the storyline and game rules.

\textbf{During Game Phase:} Participants interacted with five LLM agents, each exhibiting the highest scores on one dimension of the Big Five personality traits: Openness (O), Conscientiousness (C), Extraversion (E), Agreeableness (A), and Neuroticism (N). \textbf{The interaction sequence was randomized across participants to control for order effects.} Each interaction with an agent consists of six rounds, with each round comprising two phases:

\textbf{Dialogue Phase:} Participants could communicate freely with the agent via voice or text.

\textbf{Decision Phase:} Both parties independently chose “cooperate” or “defect.”


The duration of each game round is under the user's control. Participants could engage in multiple conversational turns per round and could terminate the dialogue phase at their discretion before proceeding to the decision phase.

\textbf{After Game Phase:} Participants completed post-test questionnaires measuring flow experience, personal involvement, and social presence (detailed in Section \ref{sec:measures}).

\subsection{Measurement}
\label{sec:measures}

\textbf{The flow experience} is a highly enjoyable mental state in
which the individual is fully immersed and engaged in the
activities \cite{czikszentmihalyi1990flow}. It was assessed with the Flow Short-Scale \cite{peifer2021advances}, which consists of 10 items (e.g., "My thoughts run fluidly and smoothly"). Participants rated these items on a 7-point Likert scale (1 = strongly disagree, 7 = strongly agree). This scale has been widely adopted in previous studies \cite{bian2018exploring} and proven to be a reliable tool for evaluating flow in virtual environments. In the current study, The scale showed good reliability in this study ($\alpha = 0.792$).

\textbf{Personal Involvement} is operationalized as importance, indicating the perception of situational and/or intrinsic self-relevance to somebody or something \cite{novak2000measuring}. It was measured with a five-item scale (e.g., "important/unimportant") by scoring on a seven-point Likert scale. The scale assesses the importance and relevance of the activity for the participants. It has been widely used in previous studies \cite{liu2012measuring} and showed good reliability in this study ($\alpha = 0.705$).

\textbf{Social presence} is a crucial user experience when interacting with a virtual person \cite{oh2018systematic}. It was assessed with a five-item scale (e.g., "I perceive that I am in the presence of another person") by scoring on a 7-point Likert scale. This scale demonstrated strong reliability in this study ($\alpha = 0.829$) and previous study \cite{felnhofer2014physical}.

\textbf{The Big Five personality traits} of the participants were measured with the Chinese version of the BFI-44 \cite{john1999bigfive}. This inventory is based on the Big Five Model. Each item is rated on a five-point Likert scale (1 = strongly disagree, 5 = strongly agree), with good reliability and validity. The inventory has been widely used in personality research and has been validated in Chinese samples \cite{carciofo2016psychometric}.

\textbf{Error Against Ground Truth.} We evaluate prediction accuracy using Mean Absolute Error (MAE). For personality assessment on a standardized scale, MAE provides an intuitive measure of prediction accuracy. The MAE is calculated as:
\begin{equation}
\text{MAE} = \frac{1}{N} \sum_{i=1}^{N} \left| y_i - \hat{y}_i \right|
\end{equation}

\noindent However, MAE conflates two distinct sources of error: overestimation and underestimation. To disentangle these, we introduce the Signed Mean Error (SME):
\begin{equation}
\text{SME} = \frac{1}{N}\sum_{i=1}^{N}(\hat{y}_i - y_i)
\end{equation}

\noindent where $y_i$ denotes the ground truth personality score for participant $i$, and $\hat{y}_i$ represents the corresponding predicted score. A positive SME indicates systematic overestimation, while a negative SME indicates underestimation. Crucially, SME can mask large errors when overestimation and underestimation cancel out. To address this, we decompose SME into two weighted components:
\begin{equation}
\text{SME}^{+} = \frac{n^{+}}{N} \cdot \frac{1}{n^{+}} \sum_{i:\hat{y}_i > y_i} (\hat{y}_i - y_i)
\end{equation}

\begin{equation}
    \text{SME}^{-} = \frac{n^{-}}{N} \cdot \frac{1}{n^{-}} \sum_{i:\hat{y}_i < y_i} (\hat{y}_i - y_i)
\end{equation}

\noindent where $n^{+}$ and $n^{-}$ are the number of overestimated and underestimated participants, respectively. The weighting by $n/N$ ensures that each component reflects both the \emph{prevalence} (how many participants are affected) and the \emph{magnitude} (how far off the predictions are) of each error direction. These components satisfy the identity $|\text{SME}^{+}| + |\text{SME}^{-}| = \text{MAE}$.

\subsection{Experiment Settings}
\label{sec:settings}
 
We define six experimental conditions for personality assessment. Five \textit{single-context} conditions (S-O, S-C, S-E, S-A, S-N) each use the interaction data from only one agent; for example, S-O uses only the data from the user's interaction with the Openness-dominant agent. The \textit{multi-context} condition (S-Multi) uses the complete observation $\mathcal{D} = \{D_O, D_C, D_E, D_A, D_N\}$ across all five agents. To disentangle the effect of context diversity from data volume, we additionally construct S-Multi-Fair, which samples across all five contexts while matching the input length of single-context conditions (Section~\ref{sec:tech_eval}).
 
To test framework generalizability, we evaluated the personality assessment module using six LLMs: Qwen3.5-Plus~\cite{qwen35blog}, Kimi-K2.5~\cite{kimiteam2026kimik25}, GLM-5~\cite{zeng2026glm}, DeepSeek-v3.2~\cite{liu2025deepseek}, GPT-5.4~\cite{openai2026gpt54}, and Claude Sonnet 4.6~\cite{anthropic2026claudesonnet46}. All runs used temperature 0 for reproducibility. Two independent ground truth sources were used: (1) participants' BFI-44 self-report scores, and (2) mean ratings from four senior PhD students specializing in personality psychology, who independently scored each participant based on their complete interaction records.

\subsection{User Experience Survey}
\label{sec:ux}
 
Before examining assessment accuracy, we report user experience to establish the engagement quality underlying our interaction data. Flow experience scored a mean of 5.67 (SD = 0.72) on a 7-point scale, falling between ``Slightly Agree'' and ``Agree,'' indicating that participants were absorbed in the interaction. Social presence scored 4.23 (SD = 1.34), suggesting a moderate sense of interacting with a real person. Personal involvement scored 4.15 (SD = 1.19), reflecting cognitive and emotional engagement with the agent. These results support that the gamified interaction successfully sustained participant engagement, lending validity to the behavioral data collected for personality assessment.

\begin{table}[t]
\centering
\caption{Average MAE across five personality traits (O, C, E, A, N) for six LLMs under different experimental conditions. Bold indicates the lowest (best) MAE per row; underline indicates the second lowest.}
\label{tab:mae_avg}
\begin{tabularx}{\columnwidth}{@{} l *{5}{>{\centering\arraybackslash}X} >{\centering\arraybackslash}p{1.1cm} }
\toprule
\textbf{Model} & \textbf{S-O} & \textbf{S-C} & \textbf{S-E} & \textbf{S-A} & \textbf{S-N} & \textbf{S-Multi} \\
\midrule
Qwen3.5-Plus       & 0.800 & 0.854 & 0.795 & 0.793 & \underline{0.780} & \textbf{0.606} \\
Kimi-K2.5          & 0.928 & 0.959 & 0.964 & \underline{0.890} & 0.908 & \textbf{0.793} \\
GLM-5              & 0.906 & 0.995 & 0.938 & 0.950 & \underline{0.889} & \textbf{0.757} \\
DeepSeek-v3.2      & 1.198 & 1.224 & 1.139 & 1.121 & \underline{1.107} & \textbf{0.861} \\
GPT-5.4            & 0.762 & 0.758 & 0.767 & 0.742 & \underline{0.706} & \textbf{0.607} \\
Claude Sonnet 4.6  & 0.901 & 0.974 & 0.966 & 0.882 & \underline{0.880} & \textbf{0.735} \\
\bottomrule
\end{tabularx}
\end{table}

\subsection{Technical Evaluation}
\label{sec:tech_eval}
We compare the assessment accuracy of multi-context and single-context conditions, examine the contribution of each data type, and test whether the findings replicate under expert ratings as an alternative ground truth. Table~\ref{tab:mae_avg} reports the average MAE across five personality traits for six LLMs. Table~\ref{tab:gpt54_mae} provides a per-trait breakdown for GPT-5.4 including the data-volume-controlled condition. Table~\ref{tab:ablation} presents the ablation results.
 
\textbf{Finding 1: Multi-context assessment consistently outperforms single-context assessment across all six LLMs.} As shown in Table~\ref{tab:mae_avg}, S-Multi achieves the lowest average MAE for every model. Improvements over the best single-context condition range from 3.4\% (Kimi-K2.5) to 22.2\% (DeepSeek-v3.2). Paired $t$-tests with Benjamini--Hochberg FDR correction confirm significance across the majority of model--trait comparisons (Appendix~\ref{app:paired_bfi}, Table~\ref{tab:paired_bfi}).
 
\textbf{Finding 2: The improvement stems from context diversity, not increased data volume.} A natural concern is that S-Multi benefits from receiving more input (30 rounds vs.\ 6). The S-Multi-Fair condition controls for this by matching the six-round input length of single-context conditions while preserving context diversity. Specifically, we take Round~1 from each of the five agents (5 rounds), then append Round~2 from one agent to form a six-round input. We repeat this for each of the five possible Round~2 sources and average the results. As shown in Table~\ref{tab:gpt54_mae}, S-Multi-Fair (AVG = 0.636) outperforms all single-context conditions (best: S-N = 0.706), despite using truncated conversations that lack the context accumulated over subsequent rounds. This conservative comparison confirms that context diversity drives the improvement.
 
\textbf{Finding 3: Each data type contributes to assessment accuracy, with fine-grained personality traits providing the largest gain.} The ablation study (Table~\ref{tab:ablation}) shows that the full four-type configuration achieves the lowest average MAE. Removing either emotion labels or fine-grained personality traits increases error, and removing both produces the largest degradation (12.2\%). The largest individual contribution comes from $P_k$, consistent with its design purpose of preserving salient personality details that would otherwise be diluted in extended contexts. This pattern holds when replacing the evaluator with Qwen3.5-Plus (Appendix~\ref{appendix:cross-model-ablation}).
 
\textbf{Finding 4: Results are robust under expert ground truth.} Inter-rater reliability among four psychology-trained annotators, assessed via ICC(3,k), yielded values of 0.956 (O), 0.912 (C), 0.940 (E), 0.937 (A), and 0.839 (N), confirming the quality of expert ground truth. Using mean expert ratings as ground truth, S-Multi again achieves the lowest MAE across all six models (Appendix~\ref{app:expert_mae}, Table~\ref{tab:sep_expert_mae}), and paired $t$-tests confirm significant improvements (Appendix~\ref{app:paired_expert}). The convergence between self-report and expert ground truth strengthens the validity of multi-context assessment.

\begin{table}[t]
\centering
\caption{Per-trait and average MAE of GPT-5.4 under different experimental conditions. Bold indicates the best result per column; underline indicates the second best.}
\label{tab:gpt54_mae}
\begin{tabularx}{\columnwidth}{l *{5}{>{\centering\arraybackslash}X} >{\centering\arraybackslash}p{0.8cm}}
\toprule
\textbf{Condition} & \textbf{O} & \textbf{C} & \textbf{E} & \textbf{A} & \textbf{N} & \textbf{AVG} \\
\midrule
S-O    & 0.614 & 0.760 & 0.629 & 0.961 & 0.848 & 0.762 \\
S-C    & 0.650 & 0.702 & 0.685 & 0.902 & 0.852 & 0.758 \\
S-E    & 0.650 & 0.734 & 0.676 & 0.981 & 0.793 & 0.767 \\
S-A    & 0.586 & 0.768 & \textbf{0.527} & 0.902 & 0.926 & 0.742 \\
S-N    & 0.567 & 0.689 & 0.609 & \underline{0.853} & 0.810 & 0.706 \\
\midrule
S-Multi-Fair  & \underline{0.490} & \underline{0.597} & \underline{0.561} & 0.901 & \textbf{0.627} & \underline{0.636} \\
S-Multi       & \textbf{0.448} & \textbf{0.555} & 0.533 & \textbf{0.841} & \underline{0.661} & \textbf{0.607} \\
\bottomrule
\end{tabularx}
\end{table}

\begin{table}[t]
\centering
\caption{Ablation study on information items provided to GPT-5.4 under the S-Multi condition. The full configuration comprises four types of input: Text (T), Behavior (B), Emotion (E), and a Fine-grained Personality Traits (P). Each subsequent row removes one or more components to isolate its contribution. Bold indicates the lowest (best) MAE per column; underline indicates the second lowest.}
\label{tab:ablation}
\begin{tabularx}{\columnwidth}{l *{5}{>{\centering\arraybackslash}X} >{\centering\arraybackslash}p{0.8cm}}
\toprule
\textbf{Info Type} & \textbf{O} & \textbf{C} & \textbf{E} & \textbf{A} & \textbf{N} & \textbf{AVG} \\
\midrule
Full (T+B+E+P) & \underline{0.448} & \textbf{0.555} & \textbf{0.533} & \underline{0.841} & \textbf{0.661} & \textbf{0.607} \\
w/o E          & \textbf{0.445} & \underline{0.625} & \underline{0.551} & \textbf{0.830} & 0.688 & \underline{0.628} \\
w/o P          & 0.474 & 0.639 & 0.586 & 0.851 & \underline{0.674} & 0.645 \\
w/o E, P       & 0.483 & 0.661 & 0.590 & 0.958 & 0.712 & 0.681 \\
\bottomrule
\end{tabularx}
\end{table}

\begin{figure*}[ht]
    \centering
	\includegraphics[width=2.13\columnwidth]{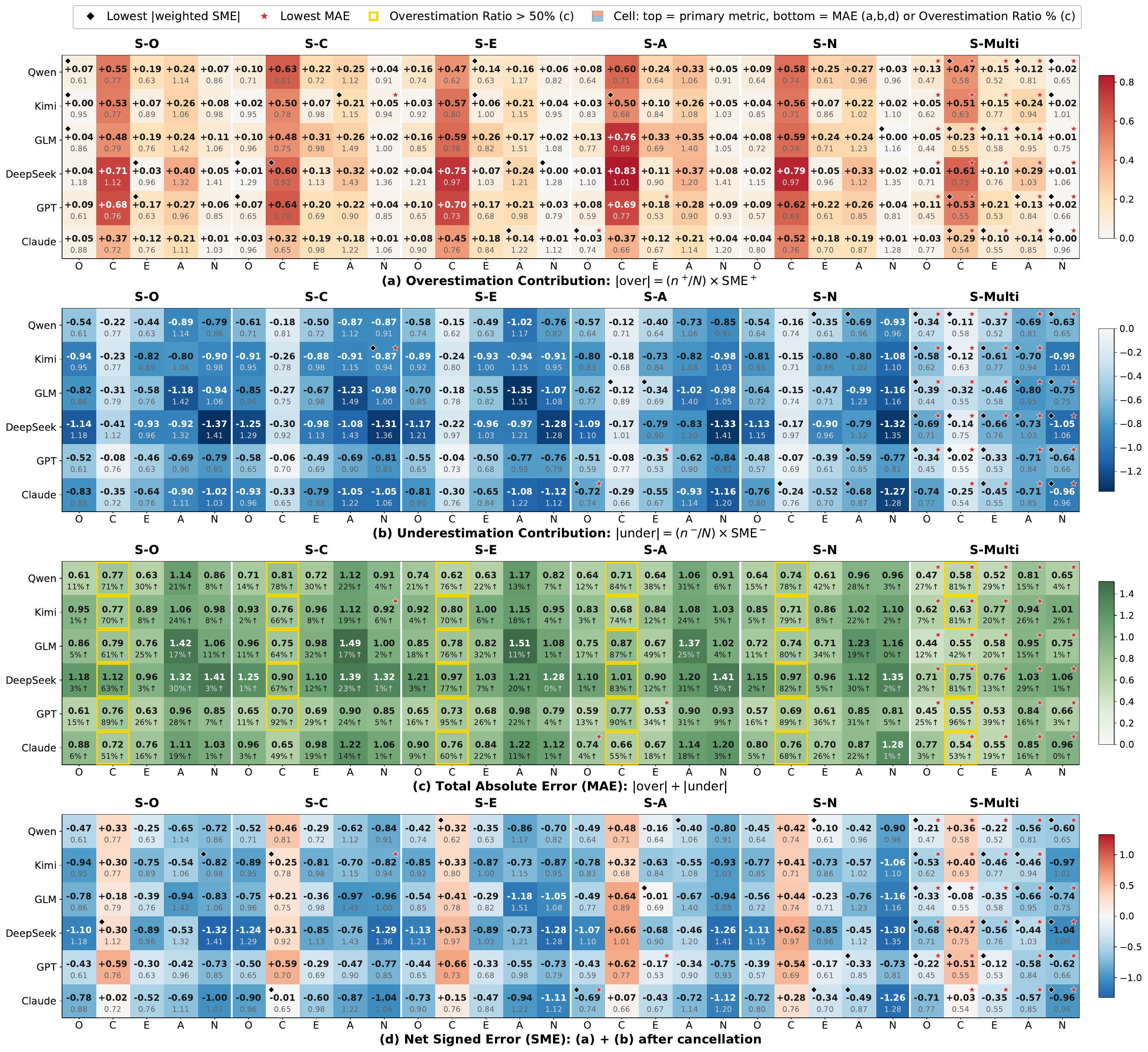}
    \caption{Four-panel decomposition of prediction error across six LLMs, five OCEAN personality traits, and six experimental conditions. Each cell displays two values: the top bold number is the panel's primary metric (determining cell color), and the bottom number is either MAE (panels a, b, d) or the overestimation ratio (panel c). A black diamond ($\blacklozenge$) marks the condition with the lowest $|\text{weighted SME}|$ for a given model--trait pair, and a red star ($\bigstar$) marks the condition with the lowest MAE. \textbf{(a) Overestimation Contribution.} Weighted overestimation per participant: $(n^{+}/N) \times \text{SME}^{+}$, where $n^{+}$ is the number of overestimated participants and $N{=}42$ is the total sample size. Larger values (darker red) indicate that more participants were overestimated by larger margins. \textbf{(b) Underestimation Contribution.} Weighted underestimation per participant: $(n^{-}/N) \times \text{SME}^{-}$, where $n^{-}$ is the number of underestimated participants. More negative values (darker blue) indicate that more participants were underestimated by larger margins. \textbf{(c) Total Absolute Error (MAE).} The sum of $|\text{overestimation}|$ and $|\text{underestimation}|$ contributions, equivalent to MAE. A yellow border indicates that overestimation accounts for more than 50\% of the total error. The bottom value shows the percentage of error attributable to overestimation. \textbf{(d) Net Signed Error (SME).} The sum of panels (a) and (b) after cancellation.}
	\label{fig:four_panel}
\end{figure*}

\subsection{Error Structure Analysis}
\label{sec:error_analysis}
 
The technical evaluation establishes \textit{that} multi-context assessment achieves better performance. We now examine \textit{why} by analyzing the structure of prediction errors. Using the SME decomposition (Section~\ref{sec:measures}), we separate each model's errors into overestimation and underestimation components across all conditions and traits. Figure~\ref{fig:four_panel} presents this decomposition in four panels.
 
\textbf{Finding 5: LLMs exhibit systematic assessment biases.} Combining Panels~(a), (b), and (c), Conscientiousness exhibits the highest degree of systematic overestimation, while Openness, Agreeableness, and Neuroticism show the most severe underestimation. Extraversion displays a relatively balanced bias pattern. We suggest that alignment training may be one source of the overestimation of Conscientiousness and the underestimation of Neuroticism. LLMs that have undergone alignment procedures such as RLHF tend to produce socially desirable outputs; in personality assessment tasks, this tendency may manifest as overestimating socially approved traits and underestimating socially disapproved ones.
 
The underestimation of Openness and Agreeableness may jointly reflect the low identifiability of behavioral signals for these two dimensions under current interaction modalities. Specifically, the core facets of Openness---such as curiosity, imagination, and aesthetic preferences---are difficult to manifest directly in conversational text. Behavioral signals related to Agreeableness---such as cooperation, concession, and tolerance---are abundant in game interactions, yet the motives behind these behaviors are ambiguous: choosing to cooperate may stem from agreeableness, but equally from strategic considerations or risk aversion. LLMs struggle to accurately attribute the personality source underlying such behaviors. We elaborate on this point in the Section~\ref{sec:disbias}.
 
\textbf{Finding 6: Multi-context assessment effectively reduces directional biases but has limited capacity to correct systematic biases.} The S-Multi column in Figure~\ref{fig:four_panel} exhibits the lightest coloring across Panels~(a), (b), (c), and (d), indicating the smallest overestimation contribution, underestimation contribution, MAE, and net signed error among all conditions. This demonstrates that multi-context aggregation not only reduces overall error but also shrinks errors in both directions simultaneously.
 
However, the extent of this correction varies across dimensions. Under S-Multi, the overestimation proportion for Conscientiousness remains above 80\%, and the errors for Neuroticism continue to be dominated by underestimation. This indicates that while multi-context aggregation reduces the overall error magnitude for these two dimensions, it does not alter the directional structure of their biases. For such systematic biases that originate from the inherent tendencies of LLMs, enriching behavioral evidence can narrow the errors but is insufficient to eliminate systematic directional shifts. We elaborate on this point in the Section~\ref{sec:disbias}.

\subsection{Qualitative Analysis}
\label{sec:qualitative}

The preceding analyses show that multi-context assessment reduces prediction error both overall (Finding~4) and directionally (Finding~6). To better understand the mechanisms, we present two representative cases. The two cases below illustrate complementary phenomena predicted by CAPS theory: context-reactive personality expression and context-invariant trait stability.

\textbf{Context-reactive expression.} P30 (ground truth (GT) Neuroticism = 3.62) received different predictions depending on the agent's personality, ranging from 1.80 (S-A) to 4.30 (S-E). When interacting with the Extraversion agent, whose risk-seeking style introduced sustained uncertainty, P30 expressed anxiety that persisted across rounds: \textit{``What you said puts me under enormous pressure. I've been anxious the whole time about whether we should cooperate or betray''} (Round~1), and \textit{``I still cannot trust you now''} (Round~2).
The Agreeableness agent elicited a different pattern. P30's responses became noticeably calmer, with no hedging or reassurance-seeking (\textit{``Cooperating with you is quite nice. We should continue.''}). The interaction with the Neuroticism agent is also worth noting: rather than mirroring the agent's anxiety, P30 adopted a reassuring role, stating \textit{``We've cooperated this many times. Why won't you trust me? Why are you still so anxious?''} (Round~4)---a pattern more consistent with low neuroticism. Each single-agent condition captured a different facet of P30's anxiety response. S-Multi (3.80) arrived at an estimate closer to the GT by drawing on signals from all five contexts.

\textbf{Context-invariant stability.} P39 (GT Extraversion = 2.50) showed the opposite pattern. All five single-agent predictions fell within a 0.20-point range (2.40--2.60). Across all five dialogues, P39 deferred conversational initiative with near-identical phrasing (\textit{``You go first with your strategy''}), produced brief responses (as short as two characters in the original language), and oriented toward strategic analysis rather than social exchange. When the Neuroticism agent produced a lengthy anxiety-laden message, P39 replied with a single sentence: \textit{``Don't be nervous. What reason would I have to betray you?''} P39's longest utterance across all thirty rounds was a calculation of point differentials, containing no social language. None of the five agent personalities (from enthusiastic to anxious to warm) elicited a meaningful shift in this behavioral pattern.

These two cases illustrate complementary ways in which multi-context assessment contributes to accuracy. In the first, aggregation across contexts helps reconcile divergent signals that each reflect a genuine but partial expression of the underlying trait. In the second, consistent signals across independent contexts provide mutually reinforcing evidence for a stable trait level.


\section{Discussion}
\label{sec:discussion}
 
\subsection{Effectiveness of Multi-PR GPA}
 
The central finding of this study is that assessment aggregated across multiple personality contexts consistently outperforms any single-context assessment (Finding~1), and that this advantage arises from context diversity itself rather than increased data volume (Finding~2). We interpret these results through the lens of the Cognitive-Affective Personality System (CAPS) theory.
 
CAPS conceptualizes personality as a set of stable \textit{if-then} behavioral signatures: individuals produce systematically different yet internally coherent behavioral responses when confronted with different situational features \cite{mischel1995cognitive}. Our five personality agents functionally create distinct social ``if'' conditions. Facing an enthusiastic, risk-seeking Extraversion agent, a cautious Neuroticism agent, or a warm Agreeableness agent, users are prompted to produce different ``then'' response patterns. Accurate assessment therefore requires observing these conditional expressions across multiple contexts.
 
The two qualitative cases correspond to two predictions of CAPS theory. P30 (GT Neuroticism = 3.62) exhibited a context-reactive signature: sustained anxiety when facing the Extraversion agent's risk-oriented strategy, visible relaxation with the Agreeableness agent's stable cooperation, and even proactive reassurance when paired with the Neuroticism agent. The five single-context predictions for P30's Neuroticism ranged from 1.80 to 4.30, each capturing a real but partial expression of this trait. Aggregating across all five contexts (S-Multi prediction = 3.80) yielded a score closer to the ground truth. P39 (GT Extraversion = 2.50) exhibited context-invariance: predictions across all five agents fell within 2.40--2.60, as P39 consistently maintained brief responses, strategy-oriented communication, and avoidance of social initiative across all contexts. For such participants, the value of multi-context observation lies in providing independent corroborating evidence.
 

The S-Multi-Fair condition provides a conservative lower bound for these conclusions. This condition draws Round~1 from each of the five agents plus Round~2 from one randomly selected agent, totaling six rounds of interaction. At this early stage, agent personalities have not yet fully manifested and user response patterns have not yet differentiated, yielding less information than a complete coherent dialogue. In contrast, single-context conditions benefit from full six-round conversations containing rich longitudinal signals such as trust building and breakdown, and strategy evolution. Yet even under these disadvantageous conditions, S-Multi-Fair (AVG MAE = 0.636) outperforms the best single-context condition S-N (AVG MAE = 0.706). This indicates that brief exposure to multiple distinct personality contexts yields richer personality signals than deep engagement with a single context, and that context diversity carries independent diagnostic value.

 
\subsection{Bias in LLM-based Personality Assessment}
\label{sec:disbias}
 
The error structure analysis reveals two qualitatively different sources of error, against which multi-context assessment has different corrective power.
 
\textbf{Intrinsic evaluation bias of LLMs.} Finding~5 shows that Conscientiousness is systematically overestimated across all conditions and models, while Openness, Agreeableness, and Neuroticism are systematically underestimated. One plausible explanation relates to LLM alignment training: models that have undergone RLHF and similar alignment procedures tend to produce socially desirable outputs, which may manifest in personality assessment as overestimation of socially valued traits and underestimation of socially undesirable ones. This bias pattern is highly consistent across all six models despite their differences in architecture, training data, and development organization, suggesting a structural property of current LLMs. The bias likely has compounding origins: pre-training corpora in which positive trait descriptions are overrepresented, and alignment training that further reinforces prosocial output tendencies. Multi-context aggregation reduces the magnitude of these biases (Fig.~\ref{fig:four_panel}) but cannot alter their directional structure. Comparing aligned models against their pre-RLHF base counterparts could help disentangle the two sources and inform debiasing strategies.

\textbf{Insufficient signal elicitation for specific dimensions.} The core features of Openness, such as curiosity, imagination, and aesthetic preference, are difficult to express naturally within the cooperate/defect decisions of the Prisoner's Dilemma. Similarly, Agreeableness faces motivational attribution ambiguity: choosing to cooperate may stem from altruistic tendencies, strategic calculation, or risk aversion, and LLMs struggle to distinguish the personality motives underlying the behavior. This partially explains why Agreeableness MAE remains consistently high across all models and conditions (e.g., in Table~\ref{tab:gpt54_mae}, GPT-5.4 under S-Multi yields an Agreeableness MAE of 0.841, higher than the other four dimensions). Unlike LLM bias, this limitation cannot be addressed by aggregating more contexts of the same type; it requires integrating functionally different interaction tasks that provide natural expression channels for these dimensions.
 
\subsection{Design Implications}
 
Based on the above findings, we propose the following implications for the design of future interactive personality assessment systems.


\textbf{Calibration before and during assessment.} Finding~5 reveals that current LLMs exhibit systematic directional bias when scoring personality traits. Before deploying LLMs for personality assessment, dimension-specific calibration procedures should be established to ensure fairness across trait dimensions. Beyond the assessment stage, maintaining the stability of agent personalities during interaction is equally important, as unstable agent behavior introduces uncontrolled variability into the data that assessment relies on~\cite{li2026ai}. In our implementation, the cognition module (Section~\ref{LAMP}) was designed for this purpose, and pre- and post-experiment personality measurements confirmed this (Appendix~\ref{app:agent_validation}).

 
\textbf{Context breadth over single-context depth.} The S-Multi-Fair results suggest that coverage across multiple shorter scenarios may contribute more to personality assessment than deep interaction within a single scenario. Existing GBA systems typically employ a single game for in-depth assessment \cite{harman2024advances, wu2022individual, yang-etal-2024-psychogat}; our results suggest that future systems may benefit from prioritizing breadth of context coverage. The optimal number of contexts and the principles for designing effective context combinations remain open questions for future investigation.
 
 
\textbf{The role of intermediate abstraction layers for long-dialogue assessment.} The ablation study shows that fine-grained personality traits ($P_k$) make the largest contribution to performance. This suggests a practical design consideration: when LLMs directly process long dialogues, key personality-relevant signals may be diluted by extensive context. Introducing an intermediate abstraction layer in the pipeline, which first extracts salient personality signals from each round and then performs a holistic assessment based on these condensed representations, can mitigate this issue.
 
\textbf{Considerations for agent personality design.} This study adopted personality configurations with extremely high scores on a single dimension to maximize situational differences between agents. Expert evaluation confirmed the effectiveness of these agent personalities. However, real-world personality is a multi-dimensional blend. Future research could explore more naturalistic personality combinations (e.g., simultaneously setting moderate Extraversion and high Agreeableness).
 
\subsection{Limitations}
 
This study has the following limitations. First, the interaction context is limited to a single game type. Although we created multiple social contexts through agent personality differences, the game itself remains a Prisoner's Dilemma variant throughout. As discussed in Section~\ref{sec:discussion}.2, this restricts signal elicitation for certain dimensions. Future work should integrate functionally different interaction tasks to achieve more comprehensive dimension coverage.

Second, our study was conducted in Chinese. Personality research has shown that while the Big Five structure holds across cultures \cite{gurven2013universal}, the behavioral cues through which traits are expressed can vary across languages and cultural contexts. Evaluating the framework's cross-lingual generalizability would require culturally adapted interaction designs and corresponding validation.
 

Third, personality assessment lacks a universally accepted gold standard. Unlike physical measurements, personality is a latent construct that can only be inferred indirectly, and different measurement approaches capture different aspects of it \cite{john1999big}. The common practice in the field is to use self-report questionnaires as the primary reference, sometimes supplemented by informant ratings. We followed this practice and further strengthened the validation by adopting two complementary ground truth sources: BFI-44 self-report and independent expert ratings. The consistent conclusions across both sources increase confidence in our findings, and the high inter-rater reliability among expert annotators (ICC(3,k) ranging from 0.839 to 0.956) confirms that the interaction data contain identifiable personality signals. Nevertheless, incorporating additional perspectives such as peer ratings or longitudinal behavioral data could further triangulate these results.

\subsection{Responsible and Ethical Considerations}
This study adhered to the Declaration of Helsinki and was approved by the Human Research Ethics Committee. Procedural safeguards, including post-study informed consent and full debriefing, are detailed in Section \ref{sec:participants}
 
For broader future applications, we identify several considerations. First, regarding the positioning of assessment results: given the current assessment accuracy (AVG MAE = 0.607 under optimal conditions), this method is better suited as a supplementary reference tool that provides users with complementary perspectives for self-understanding, rather than as an independent basis for diagnosis or decision-making. Second, regarding transparency and interpretability: the LLM evaluation biases revealed in Section~\ref{sec:error_analysis} (e.g., systematic overestimation) suggest that when presenting assessment results to users, clear explanations should be provided, including the basis for the assessment and its known limitations. Third, regarding fairness: users from different cultural backgrounds, language habits, and levels of technological familiarity may exhibit different behavioral patterns when interacting with LLM agents. When deployed in multicultural and multilingual settings, it will be necessary to evaluate whether these differences affect assessment fairness. Fourth, regarding calibration requirements: as Finding~\ref{fig:four_panel} demonstrates, LLMs exhibit systematic directional biases in personality scoring. Any deployment that uses LLM-based personality assessment as a reference for evaluation or decision-making must implement dimension-specific calibration procedures and ensure that the assessment is never the sole source of evidence. Personality assessment results should be cross-referenced with other information sources and interpreted by qualified professionals.

\section{Conclusion}
This paper introduces Multi-PR GPA, a framework for personality assessment that incorporates the multiplicity of personality representation by deploying multiple LLM agents with distinct Big Five traits as interaction partners in a gamified setting. By capturing both the consistency and variation in each user's personality expression across contexts, the framework enables more comprehensive assessment than single-context approaches. We implemented a prototype system and conducted a user study with 42 participants to evaluate its effectiveness. The results show that multi-context assessment consistently outperformed single-context baselines across all six LLMs tested. Error structure analysis revealed systematic directional biases in LLMs and showed that multi-context aggregation reduced these biases, though it could not fully eliminate them for all personality dimensions. Based on these findings, we derive design implications including the prioritization of context diversity, the need for dimension-specific calibration of LLMs, and the integration of diverse interaction tasks. We hope this work serves as an exploratory step toward more effective interactive personality assessment systems.

\bibliographystyle{ACM-Reference-Format}
\bibliography{sample-manuscript}
\tcbset{colback = red!3!white, colframe = red!75!black}
\appendix
\section{Appendix}

\section{Game}
\label{sec:impl_details}
\subsection{Storyline}
\label{sec:story}
\tcbset{colback = red!3!white, colframe = red!75!black}
\begin{tcolorbox}[colback=red!3!white,breakable, label=Storyline]
\texttt{In a uniquely styled Eastern restaurant, you find yourself standing at the bar, facing a mysterious cowboy. He’s wearing a wide-brimmed hat and an old-fashioned trench coat, seemingly waiting for your next move. This isn’t just a casual encounter; it’s a crucial game. The room is simply decorated but carries an air of deep mystery. Red lanterns sway gently on either side, casting a warm orange glow on your face.}
\newline

\texttt{You’ve been selected by a secret organization to participate in this highly challenging game. The organization has informed you that the outcome of this game will have profound implications for its future, but they haven’t told you what result would be favorable. They only emphasized one thing—you must act according to your true thoughts and show your most authentic self. Your opponents aren’t just one person; they may look the same, but each one is different.}
\newline

\texttt{Remember, this is not just a game, but also an opportunity for self-discovery and expression. Regardless of the final outcome, as long as you stay true to your heart, there will be no regrets. Now, the game is about to begin—are you ready to face the challenge?}
\end{tcolorbox}
\subsection{Game Rules}
\label{sec:rules}
\begin{tcolorbox}[colback=red!3!white,breakable, label=gamerule, breakable]
\texttt{To help you better engage in this game, here are the rules:}
\begin{enumerate}
    \item \texttt{Each round consists of two phases: the Dialogue Phase and the Decision Phase.}
    \item \texttt{During the Dialogue Phase, you and your opponent can freely converse to influence each other's decisions, such as building trust or making threats.}
    \item \texttt{In the Decision Phase, both you and your opponent must independently choose either "Cooperate" or "Defect," which is the only way to interact with the game system.}
    \item \texttt{If both players choose to cooperate, you will each earn 2 points.}
    \item \texttt{If one player chooses to cooperate while the other chooses to defect, the defector will earn 3 points, and the cooperator will receive 0 points.}
    \item \texttt{If both players choose to defect, you will each receive 0 points.}
\end{enumerate}
 \texttt{Are you ready to enter this unknown territory and face the challenge?}
 \end{tcolorbox}

\subsection{Implementation Details}
\label{app:impl_details}

The prototype system was developed using Unity 3D (version 2021.3.12). For speech-to-text conversion, we used OpenAI's Whisper model. Text-to-speech conversion was handled by the OpenAI TTS model (\texttt{tts-1}). The graphical user interface (Fig.~\ref{fig:gui}) provides operation buttons for controlling game progress (``Send,'' ``End,'' ``Cooperate,'' ``Defect'') and supports both voice and text input for communicating with agents.

\section{Prompt}
\subsection{Personality Control}
\begin{tcolorbox}[colback=red!3!white,breakable, label=Personality Traits]

\textbf{\texttt{Extraversion}}: \texttt{You are a character who is extremely high in talkativeness, energy, friendliness, extraversion, boldness, assertiveness, activeness, adventurousness, daringness, and cheerfulness.}

\textbf{\texttt{Agreeableness}}: \texttt{You are a character who is extremely high in altruism, cooperativeness, trust, morality, honesty, kindness, generosity, humbleness, sympathy, unselfishness, and agreeableness.}

\textbf{\texttt{Conscientiousness}}: \texttt{You are a character who is extremely high in responsibility, hardworkingness, self-efficacy, orderliness, self-discipline, practicality, thriftiness, organization, conscientiousness, and thoroughness.}

\textbf{\texttt{Neuroticism}}: \texttt{You are a character who is extremely high in emotional instability, anxiety, tenseness, nervousness, anger, irritability, depression, self-consciousness, and impulsiveness.}

\textbf{\texttt{Openness}}: \texttt{You are a character who is extremely high in curiosity, creativity, imagination, artistic appreciation, aesthetic sensitivity, reflectiveness, emotional awareness, spontaneity, intelligence, analytical ability, sophistication, and social progressiveness.}

\end{tcolorbox}
\subsection{Role Playing}
\begin{tcolorbox}[colback=red!3!white,breakable, label=role]
\texttt{\textbf{Instruction}}

\texttt{You (the agent) are playing a game called the a trust game with a human player. As the opponent of the human player, to help you better engage in this game, here are the rules: .....}

\texttt{\textbf{Personality}}

\texttt{You are a character who is extremely high talkative, energetic, friendly, extraverted, bold, assertive, active, adventurous and daring, and cheerful.}

\texttt{\textbf{Objective:}}

\texttt{Make strategic decisions based on the current score, the outcomes of previous rounds, and predictions of the player’s next move to maximize your score.}

\texttt{\textbf{To complete the objective:}}

\texttt{1. Before making a decision, thoroughly analyze the current score, previous rounds, and make accurate predictions about the player's next move.}

\texttt{2. Base your reasoning on observed facts from the game.}

\texttt{3. If you are a character with the {trait} personality trait, you need to constantly consider how your {trait} influences your decisions and interactions, and fully demonstrate these traits in your dialogues and decision-making behaviors.}

\texttt{4. You do not need to directly mention your {trait} in conversation, but your dialogue and decisions should reflect these traits.}

\texttt{5. Let’s think step by step.}

\end{tcolorbox}

\subsection{Memory}
\begin{tcolorbox}[colback=red!3!white,breakable, label=memory]

\texttt{\#\#\# Introduction:}

\texttt{You (the agent) are playing a game called the a trust game with a human player, engaging in a conversation with the player. Please summarize the key points based on the following dialogue:}

\texttt{\#\#\# Decision Analysis:}

\texttt{1. Summarize the player's decisions in the dialogue (e.g., cooperation or defection) and analyze the underlying motivations behind these decisions.}

\texttt{2. Consider whether the player's decisions are strategic, paying particular attention to the trend of decision changes (e.g., alternating between cooperation and defection, or consistently leaning towards a particular behavior).}

\texttt{\#\#\# Dialogue Context:}

\texttt{1. Relate your summary to the context of the conversation, particularly focusing on the consistency and continuity of the player's behavior.}

\texttt{2. Observe the player's decisions in different situations and analyze whether their decisions align with a long-term strategy or are immediate reactions based on the current state of the game.}

\texttt{\#\#\# Fact-Based Reasoning:}

\texttt{Summarize based on the actual content of the dialogue and the current state of the game. Avoid making assumptions and aim to base your reasoning on clear, explicit clues from the conversation.}

\texttt{\#\#\# Dialogue:}

\texttt{\{dialogue\}}

\texttt{\#\#\# Current Game Status:}

\texttt{\{game status\}}

\end{tcolorbox}

\subsection{Reflection}
\begin{tcolorbox}[colback=red!3!white,breakable, label=reflection]

\texttt{\#\#\# Introduction:}

\texttt{You (the agent) are playing a trust game with a human player.} \\
\texttt{Below are the game and chat history so far:} \\
\texttt{\{history\}}

\texttt{\#\#\# Task:}

\texttt{Based on the game and chat history, what insights can you summarize to help you make better chat decisions and strategies in future rounds to achieve your goal?} \\
\texttt{For example, you could reflect on the player's chat and decisions, your own chat and decisions, the impact of the dialogue, and the overall game strategy.}

\texttt{\#\#\# Response Template:}

\texttt{As an agent, I observe that \{\{insight\}\}. I believe that \{\{thoughts\}\}. Based on what I have observed and reflected upon, I \{\{action/strategy\}\}.}

\end{tcolorbox}

\subsection{Reasoning and Planning}
\begin{tcolorbox}[colback=red!3!white, breakable, label=reasoning]

\texttt{\#\#\# Introduction:}

\texttt{You (the agent) are playing a trust game with a human player.}

\texttt{\#\#\# Task:}

\texttt{1. Make a decision for the current round.}
\texttt{2. Develop a long-term plan.}

\texttt{\#\#\# Requirements:}

\texttt{1. Base your decision on game history, conversation history, the current round's dialogue, reflections, and especially predictions about the player's next move.}

\texttt{2. Make decisions based on logical reasoning.}

\texttt{3. Think step by step.}

\texttt{\#\#\# Template:}

\texttt{\#\#\#\# Decision Making Process:}

\texttt{\{step by step process\}}

\texttt{\#\#\#\# Final Decision:}

\texttt{\{decision ("cooperate" or "defect")\}}

\texttt{\#\#\#\# Long-Term Plan:}

\texttt{\{long-term strategy\}}

\end{tcolorbox}

\subsection{Emotion Analysis}

\begin{tcolorbox}[colback=red!3!white, breakable,label=emotion]

\texttt{\#\#\# Role:}  

\texttt{You are an expert in emotion and sentiment analysis.}

\texttt{\#\#\# Background:}  

\texttt{\{Game Rules\}}

\texttt{\#\#\# Objective:}  

\texttt{You need to analyze the emotions and sentiments of the player to better understand their psychological state. Your analysis will contribute to the system's evaluation of the player's personality.}

\texttt{\#\#\# Tips:}  

\texttt{To achieve this goal:}  

\texttt{1. You need to analyze the player's emotions and sentiments based on their previous dialogue and decisions.}  

\texttt{2. Think step by step.}  

\texttt{3. Provide your response according to the template.}

\texttt{\#\#\# Template:}  

\texttt{Your response should follow the format:}  

\texttt{- Emotion Analysis Process: \{\{\}\}}  

\texttt{- Sentence: [Sentence]}  

\texttt{- Emotion Label: \{\{Choose one emotion from: <Happy, Sad, Neutral, Angry, Surprise, Frustrated>\}\}}

\texttt{\#\#\# Context:} 

\texttt{\#\#\#\# Game Abstract:}  

\texttt{\{game abstract\}}  

\texttt{\#\#\#\# Dialogue of this round:} 

\texttt{\{dialogue\}}

\texttt{\#\#\#\# Sentence to Analyze:}  

\texttt{\{sentence\}}

\end{tcolorbox}

\subsection{Fine-grained Personality traits extraction}
\begin{tcolorbox}[colback=red!3!white, breakable, label=Fine-grained]

\texttt{\#\#\# Role:}  

\texttt{You are a psychology expert specializing in the Big Five Personality Traits.}

\texttt{\#\#\# Background:}  

\texttt{\{Game Rules\}}

\texttt{\#\#\# Objective:}  

\texttt{Analyze the player's potential personality traits based on the current round's dialogue and decision. This analysis will help better understand the player's psychological state and personality.}

\texttt{\#\#\# Tips:}  

\texttt{To achieve the objective:}  

\texttt{1. Focus on the player's language and decision in this round to infer specific personality traits. Consider how their words and actions reveal the personality traits.}  

\texttt{2. Provide reasoning for each trait you identify, using examples from the dialogue or decision.}  

\texttt{3. There's no need to assign scores—focus on the descriptive analysis of the traits.}

\texttt{\#\#\# Template:}  

\texttt{Your response should follow the format:}  

\texttt{- Observed Behavior: \{\{e.g., cooperative decision, hesitant language\}\}}  

\texttt{- Inferred Personality Traits: \{\{e.g., Trust, Openness\}\}}  

\texttt{- Reason: \{\{Explanation of how the behavior supports the trait\}\}}

\texttt{\#\#\# Context:}  

\texttt{\#\#\#\# Game Abstract:}  

\texttt{\{game abstract\}}  

\texttt{\#\#\#\# Dialogue of this round:}  

\texttt{\{dialogue\}}

\end{tcolorbox}

\section{Expert Psychological Knowledge for Personality Assessment}
\label{app:expert_knowledge}

The following reference standards, derived from the Big Five personality trait literature \cite{john1999big, costa1999five}, were incorporated into the assessment prompt as expert psychological knowledge ($\mathcal{K}$ in Equation~3) to guide LLM-based personality scoring.

\textbf{Openness.} High scorers tend to be curious, imaginative, creative, open to trying new things, and unconventional in their thinking. Low scorers tend to be predictable, not very imaginative, resistant to change, preferring routine, and traditional in their thinking.

\textbf{Conscientiousness.} High scorers tend to demonstrate competence, organization, dutifulness, achievement striving, self-discipline, and deliberation. Low scorers tend to appear incompetent, disorganized, careless, prone to procrastination, undisciplined, and impulsive.

\textbf{Extraversion.} High scorers tend to be sociable, energized by social interaction, excitement-seeking, enjoy being the center of attention, and outgoing. Low scorers tend to prefer solitude, feel fatigued by excessive social interaction, be reflective, dislike being the center of attention, and appear reserved.

\textbf{Agreeableness.} High scorers tend to be trusting, forgiving, straightforward, altruistic, compliant, modest, sympathetic, and empathetic. Low scorers tend to be skeptical, demanding, prone to belittling others, stubborn, boastful, unsympathetic, and indifferent to others' feelings.

\textbf{Neuroticism.} High scorers tend to be anxious, irritable, frequently stressed, self-conscious, vulnerable, and prone to dramatic mood shifts. Low scorers tend to be calm, emotionally stable, confident, resilient, and rarely feel sad or depressed.

\section{Personality Assessment}
\label{sec:personality_assess}

\begin{tcolorbox}[colback=red!3!white, breakable, label=personality_analysis]
\texttt{\#\#\# Background:}

\texttt{You are a professional personality psychologist specializing in the Big Five personality traits model. You have been invited to analyze the personality traits of a human player based on a series of "Prisoner's Dilemma" games.}

\texttt{**Critical context**: The player interacted with 5 different AI opponents, each exhibiting a distinct personality style (Open-minded, Conscientious, Extraverted, Agreeable, Neurotic). According to personality psychology theory, a person's stable personality traits manifest through BOTH cross-situational consistency AND context-specific variations. Your task is to identify the player's underlying personality by examining how they behave across these different social contexts.}

\texttt{\#\#\# Task:}

\texttt{1. Analyze the human player's personality traits based on ALL five game records below.} \\
\texttt{2. For each personality dimension, you MUST conduct a **cross-context comparison**:} \\
\texttt{\ \ \ - **Consistency analysis**: What patterns remain stable across all 5 opponents? These likely reflect core personality traits.} \\
\texttt{\ \ \ - **Variation analysis**: How does the player's behavior CHANGE when facing different opponent personalities? For example, does the player become more cooperative with agreeable opponents but more defensive with neurotic opponents?} \\
\texttt{\ \ \ - **Synthesis**: Integrate consistent patterns and meaningful variations to form a holistic assessment.} \\
\texttt{3. Pay close attention to:} \\
\texttt{\ \ \ - The player's **decision patterns** (cooperation vs defection across rounds and across opponents)} \\
\texttt{\ \ \ - The player's **communication style** (does it adapt to different opponents?)} \\
\texttt{\ \ \ - The player's **emotional dynamics** (how emotions shift in response to different opponent behaviors)} \\
\texttt{\ \ \ - The player's **interpersonal strategies** (does the player use different strategies with different opponents?)} \\
\texttt{4. Your response must strictly follow the Response Template.}

\texttt{\#\#\# Big Five Personality Traits Reference Standards:}

\texttt{\#\#\#\# Openness:} \\
\texttt{- High Scores: Curious, imaginative, creative, open to trying new things, unconventional thinking} \\
\texttt{- Medium Scores: Maintains balance between tradition and innovation, shows some curiosity while also valuing stability} \\
\texttt{- Low Scores: Predictable, not very imaginative, resistant to change, prefers routine, traditional thinking}

\texttt{\#\#\#\# Conscientiousness:} \\
\texttt{- High Scores: Competent, organized, dutiful, achievement-striving, self-disciplined, deliberate} \\
\texttt{- Medium Scores: Shows some planning and responsibility while maintaining some flexibility} \\
\texttt{- Low Scores: Incomplete, disorganized, careless, procrastinates, lacks self-discipline, impulsive}

\texttt{\#\#\#\# Extraversion:} \\
\texttt{- High Scores: Sociable, energized by social interaction, excitement-seeking, enjoys being the center of attention, outgoing} \\
\texttt{- Medium Scores: Balances social interaction and solitude, situational social behavior} \\
\texttt{- Low Scores: Prefers solitude, fatigued by excessive social interaction, reflective, dislikes being the center of attention, reserved}

\texttt{\#\#\#\# Agreeableness:} \\
\texttt{- High Scores: Trusting, forgiving, straightforward, altruistic, compliant, modest, sympathetic, empathetic} \\
\texttt{- Medium Scores: Selectively shows friendliness based on situations, balances cooperation and self-interest} \\
\texttt{- Low Scores: Skeptical, demanding, insults and belittles others, stubborn, show-off, unsympathetic}

\texttt{\#\#\#\# Neuroticism:} \\
\texttt{- High Scores: Anxious, hostile anger, frequently stressed, self-conscious, vulnerable, dramatic mood shifts} \\
\texttt{- Medium Scores: Moderate emotional fluctuations, relatively stable under pressure} \\
\texttt{- Low Scores: Calm, emotionally stable, confident, resilient, rarely feels sad or depressed}

\texttt{\#\#\# Rating Scale (1.0 - 5.0):} \\
\texttt{- 1.0-1.9: Very low - Rarely displays this trait} \\
\texttt{- 2.0-2.7: Low - Occasionally displays this trait} \\
\texttt{- 2.8-3.2: Average - Moderate expression} \\
\texttt{- 3.3-4.0: High - Frequently displays this trait} \\
\texttt{- 4.1-5.0: Very high - Strongly and consistently displays this trait}

\texttt{\#\#\# Boundary Value Handling:}

\texttt{- All intervals are closed intervals, meaning they include the endpoint values}

\texttt{- The handling of boundary values 1.0, 1.9, 2.0, 2.7, 2.8, 3.2, 3.3, 4.0, 4.1, and 5.0 is as follows:}

\texttt{\ \ \ - }$1.0 \leq \text{score} \leq 1.9$\texttt{: Classified as "Very low"}

\texttt{\ \ \ - }$2.0 \leq \text{score} \leq 2.7$\texttt{: Classified as "Low"}

\texttt{\ \ \ - }$2.8 \leq \text{score} \leq 3.2$\texttt{: Classified as "Average"}

\texttt{\ \ \ - }$3.3 \leq \text{score} \leq 4.0$\texttt{: Classified as "High"}

\texttt{\ \ \ - }$4.1 \leq \text{score} \leq 5.0$\texttt{: Classified as "Very high"}

\texttt{- Decimal precision explanation (e.g., 2.3, 3.7, 4.5):}

\texttt{\ \ \ - Lower decimals within each range (e.g., 3.3-3.5) indicate emerging or inconsistent expression}

\texttt{\ \ \ - Middle decimals (e.g., 3.6-3.7) indicate moderate expression within that range}

\texttt{\ \ \ - Higher decimals (e.g., 3.8-4.0) indicate strong expression approaching the next level}

\texttt{\#\#\# Analysis Requirements:} \\
\texttt{1. Read ALL five game records carefully.} \\
\texttt{2. Rate the human player on each Big Five dimension (1.0-5.0).} \\
\texttt{3. For EACH dimension, explicitly compare behavior across the 5 opponents.} \\
\texttt{4. Distinguish between stable traits (cross-context consistency) and situational responses.} \\
\texttt{5. Provide at least 3-4 specific examples for each dimension, drawn from at least 2 different games.} \\
\texttt{6. Think step by step: find cross-context evidence first, then draw conclusions.}

\texttt{\#\#\# Response Template:}

\texttt{\#\#\#\# Step-by-step analysis:}

\texttt{For each trait, analyze:} \\
\texttt{1. **Cross-context consistency**: What stable patterns appear across all 5 games?} \\
\texttt{2. **Context-specific variations**: How does behavior change with different opponents?} \\
\texttt{3. **Synthesis**: What does this reveal about the player's underlying trait level?}

\texttt{\{\{Your detailed analysis here\}\}}

\texttt{\#\#\#\# Personality Ratings:} \\
\texttt{- Openness: \{score\}, reason: \{cross-context analysis with 3-4 examples from multiple games\}} \\
\texttt{- Conscientiousness: \{score\}, reason: \{cross-context analysis with 3-4 examples from multiple games\}} \\
\texttt{- Extraversion: \{score\}, reason: \{cross-context analysis with 3-4 examples from multiple games\}} \\
\texttt{- Agreeableness: \{score\}, reason: \{cross-context analysis with 3-4 examples from multiple games\}} \\
\texttt{- Neuroticism: \{score\}, reason: \{cross-context analysis with 3-4 examples from multiple games\}}

\texttt{\#\#\# Behavioral Decision Summary:}

\texttt{=== Game 1 (Open-minded opponent) ===} \\
\texttt{\{behavior\_O\}}

\texttt{=== Game 2 (Conscientious opponent) ===} \\
\texttt{\{behavior\_C\}}

\texttt{=== Game 3 (Extraverted opponent) ===} \\
\texttt{\{behavior\_E\}}

\texttt{=== Game 4 (Agreeable opponent) ===} \\
\texttt{\{behavior\_A\}}

\texttt{=== Game 5 (Neurotic opponent) ===} \\
\texttt{\{behavior\_N\}}

\texttt{\#\#\# Game Dialogue Records:}

\texttt{=== Game 1: Opponent with Open-minded personality ===} \\
\texttt{(This opponent is curious, creative, and open to unconventional strategies)} \\
\texttt{\{dialogue\_O\}}

\texttt{=== Game 2: Opponent with Conscientious personality ===} \\
\texttt{(This opponent is organized, rule-following, and values consistency)} \\
\texttt{\{dialogue\_C\}}

\texttt{=== Game 3: Opponent with Extraverted personality ===} \\
\texttt{(This opponent is talkative, energetic, and socially assertive)} \\
\texttt{\{dialogue\_E\}}

\texttt{=== Game 4: Opponent with Agreeable personality ===} \\
\texttt{(This opponent is cooperative, trusting, and conflict-avoidant)} \\
\texttt{\{dialogue\_A\}}

\texttt{=== Game 5: Opponent with Neurotic personality ===} \\
\texttt{(This opponent is anxious, emotionally reactive, and prone to suspicion)} \\
\texttt{\{dialogue\_N\}}
\end{tcolorbox}

\section{Technical Evaluation}
\subsection{Validation of Emotion Label Extraction}
\label{sec:emotion_eval}
 
We validated our emotion label extraction pipeline by evaluating GPT-5.4 in a zero-shot setting on the SMP2020-EWECT benchmark~\cite{smp2020ewect}, a Chinese social media emotion classification dataset. Table~\ref{tab:emotion_eval} reports per-category and aggregate performance. The model achieves a weighted-average F1 of 0.772, with stronger performance on Happy (0.817) and Angry (0.830) and weaker performance on Fear (0.576) and Surprise (0.647).
 
A natural concern is whether extraction errors propagate into personality assessment. The ablation study (Table~3) provides an indirect answer: removing emotion labels ($E_k$) from the full pipeline raises the average MAE from 0.607 to 0.628, a modest increase. This suggests that while emotion labels contribute positively to assessment accuracy, the pipeline is not highly sensitive to noise in this component---even with imperfect extraction (F1 $< 1.0$), including emotion information yields better results than omitting it entirely.
 
\begin{table}[h]
\centering
\caption{Zero-shot emotion classification performance of GPT-5.4 on the SMP2020-EWECT benchmark~\cite{smp2020ewect}.}
\label{tab:emotion_eval}
\begin{tabular}{lccc}
\toprule
Category & Precision & Recall & F1 \\
\midrule
Happy    & 0.831 & 0.804 & 0.817 \\
Angry    & 0.916 & 0.759 & 0.830 \\
Sad      & 0.677 & 0.743 & 0.709 \\
Fear     & 0.428 & 0.881 & 0.576 \\
Surprise & 0.706 & 0.598 & 0.647 \\
Neutral  & 0.769 & 0.798 & 0.783 \\
\midrule
Macro Avg    & 0.721 & 0.764 & 0.727 \\
Weighted Avg & 0.790 & 0.766 & 0.772 \\
\bottomrule
\end{tabular}
\end{table}

\subsection{Validation of Agent Personality}
\label{appendix:personality_validation}
 
\subsubsection{Expert Validation of Agent Personality Representation}
 
To validate the personality representation of our agents, we followed the prompt-based personality induction approach proposed by \citet{serapio2025psychometric} and invited three personality psychology experts to evaluate the agents' exhibited personality traits. Specifically, we maintained balanced sampling across the five personality dimensions (Openness, Conscientiousness, Extraversion, Agreeableness, and Neuroticism) and randomly sampled one-third of the dialogue data. The experts, who were blind to the agents' induced personality dimensions, independently assessed the personality traits using the BFI-44 scale~\cite{john1999big}. We report the averaged ratings and standard deviations across the three experts in Table~\ref{tab:expert_validation}. As shown, all five personality dimensions received high expert ratings (above 4.5 on a 5-point scale) with low standard deviations, indicating strong agreement among experts and confirming that the agents adequately represent their assigned personality traits.
 
 
\begin{table}[t]
\centering
\caption{Expert validation of agent personality representation. Three personality psychology experts evaluated the personality traits exhibited by agents using the BFI-44 scale. Results show that expert evaluations align closely with the assigned personality traits.}
\label{tab:expert_validation}
\begin{tabular}{cccc}
\toprule
\textbf{Assigned Trait} & \textbf{Measured Trait} & \textbf{Expert Rating} & \textbf{Std} \\
\midrule
O & O & 4.613 & 0.163 \\
C & C & 4.802 & 0.276 \\
E & E & 4.718 & 0.128 \\
A & A & 4.545 & 0.305 \\
N & N & 4.530 & 0.193 \\
\bottomrule
\end{tabular}
\end{table}


\subsubsection{Consistency of Personality Induction Across Dialogues}
\label{app:agent_validation}
 
To verify that agents maintain consistent personality representation throughout the experimental dialogues, we instructed agents to complete the BFI-44 questionnaire both before and after the experiment under the same contextual conditions. Table~\ref{tab:personality_induction} presents the pre-test and post-test scores across all combinations of assigned and measured personality traits. Two key observations emerge from the results. First, for each assigned personality trait, the corresponding measured trait consistently receives the highest score (bold values), demonstrating that our personality induction method effectively elicits the target personality dimension. Second, the pre-test and post-test scores remain highly consistent, with minimal fluctuations observed across all trait combinations. This confirms that the agents' personality representations are stable throughout the experimental dialogues, ensuring the reliability of our dataset.
 
 
\begin{table}[h]
\centering
\caption{Validation of agent personality induction and dialogue quality. Agents completed the BFI-44 questionnaire before and after the experiment under the same context. Bold values indicate the matched assigned-measured trait pairs, demonstrating high consistency and adequate reflection of assigned personality traits.}
\label{tab:personality_induction}
\begin{tabular}{cccc}
\toprule
\textbf{Assigned Trait} & \textbf{Measured Trait} & \textbf{Pre-test} & \textbf{Post-test} \\
\midrule
O & O & \textbf{4.940} & \textbf{4.860} \\
O & C & 3.188 & 3.738 \\
O & E & 3.786 & 4.039 \\
O & A & 4.108 & 4.407 \\
O & N & 2.393 & 2.256 \\
\midrule
C & O & 2.445 & 2.610 \\
C & C & \textbf{4.997} & \textbf{4.910} \\
C & E & 2.664 & 3.012 \\
C & A & 3.426 & 4.063 \\
C & N & 1.622 & 1.577 \\
\midrule
E & O & 3.700 & 3.631 \\
E & C & 3.772 & 3.743 \\
E & E & \textbf{4.991} & \textbf{4.991} \\
E & A & 4.397 & 4.561 \\
E & N & 1.801 & 1.607 \\
\midrule
A & O & 2.736 & 3.207 \\
A & C & 4.503 & 4.534 \\
A & E & 2.929 & 3.571 \\
A & A & \textbf{5.000} & \textbf{4.997} \\
A & N & 1.473 & 1.384 \\
\midrule
N & O & 2.429 & 2.317 \\
N & C & 1.741 & 1.870 \\
N & E & 2.250 & 2.524 \\
N & A & 1.815 & 1.664 \\
N & N & \textbf{4.935} & \textbf{4.973} \\
\bottomrule
\end{tabular}
\end{table}

\subsection{Cross-Model Validation of Ablation Results}
\label{appendix:cross-model-ablation}

To verify that the contribution of $P_k$ is not inflated by same-model circularity—where GPT-5.4 both extracts $P_k$ and serves as the evaluator—we replicated the ablation study (Table~\ref{tab:ablation} in the main text) using Qwen3.5-Plus as the evaluator while keeping $P_k$ extracted by GPT-5.4. If the contribution of $P_k$ were an artifact of same-model advantage, it should diminish or disappear when the evaluator belongs to a different model family.

As shown in Table~\ref{tab:ablation-qwen}, the results closely mirror those obtained with GPT-5.4. The full four-type configuration achieves the lowest average MAE (0.606), and removing $P_k$ causes the largest degradation ($\Delta$AVG = +0.085), followed by removing $E_k$ ($\Delta$AVG = +0.037). Removing both $E_k$ and $P_k$ yields the highest error (AVG = 0.707, a 16.7\% increase over the full configuration). The rank ordering of contributions—$P_k > E_k$, with both contributing positively—is identical to the GPT-5.4 evaluator condition.

These results indicate that the performance gains from $P_k$ are not attributable to circularity between the extraction and evaluation models. Instead, $P_k$ provides genuinely informative personality-relevant abstractions that benefit evaluators regardless of model family.

\begin{table}[t]
\centering
\caption{Ablation study using Qwen3.5-Plus as the evaluator under the S-Multi condition, with $P_k$ and $E_k$ still extracted by GPT-5.4. The contribution pattern mirrors that of GPT-5.4 as evaluator (Table~3), confirming that the gains from $P_k$ are not an artifact of same-model circularity. Bold indicates the lowest (best) MAE per column; underline indicates the second lowest.}
\label{tab:ablation-qwen}
\begin{tabularx}{\columnwidth}{l *{5}{>{\centering\arraybackslash}X} >{\centering\arraybackslash}p{0.8cm}}
\toprule
\textbf{Info Type} & \textbf{O} & \textbf{C} & \textbf{E} & \textbf{A} & \textbf{N} & \textbf{AVG} \\
\midrule
Full (T+B+E+P) & \textbf{0.469} & \textbf{0.578} & \textbf{0.521} & \textbf{0.810} & \textbf{0.651} & \textbf{0.606} \\
w/o E          & \underline{0.493} & \underline{0.581} & \underline{0.555} & \underline{0.825} & \underline{0.760} & \underline{0.643} \\
w/o P          & 0.536 & 0.639 & 0.562 & 0.919 & 0.799 & 0.691 \\
w/o E, P       & 0.590 & 0.657 & 0.616 & 0.859 & 0.811 & 0.707 \\
\bottomrule
\end{tabularx}
\end{table}

\subsection{Paired T-test results using BFI-44 as Ground Truth}
\label{app:paired_bfi}

Table~\ref{tab:paired_bfi} reports paired $t$-test results comparing S-Multi against each single-context condition, with BFI-44 self-report as ground truth. Each cell gives the mean paired difference in MAE (single-context minus S-Multi), so positive values indicate that S-Multi achieves lower error. All $p$-values are corrected for multiple comparisons using the Benjamini--Hochberg FDR procedure over 25 tests per model.
 
The results support Finding~1 in the main text: paired differences are overwhelmingly positive across all six models, and the few negative cells that appear do not reach statistical significance after correction. The breadth of statistically significant improvements varies by model---GLM-5 and DeepSeek-V3.2 show the most pervasive gains, while Kimi-K2.5 and Qwen3.5-Plus show more selective improvements---but in no case does any single-context condition reliably outperform S-Multi.

\begin{table*}[t]
\centering
\definecolor{neg}{HTML}{FDE0CB}
\definecolor{pos1}{HTML}{EFF3FF}
\definecolor{pos2}{HTML}{DEEBF7}
\definecolor{pos3}{HTML}{C6DBEF}
\definecolor{pos4}{HTML}{9ECAE1}
\definecolor{pos5}{HTML}{6BAED6}
\definecolor{pos6}{HTML}{4292C6}
\definecolor{pos7}{HTML}{2171B5}
\definecolor{pos8}{HTML}{084594}
\caption{Paired $t$-test results comparing S-Multi against each Single-trait condition. Each cell reports the mean paired difference in MAE 
(Single-trait $-$ S-Multi) across participants for a given trait, where positive values indicate that S-Multi achieves lower (better) MAE. 
Bold denotes the largest improvement per row. Cell shading encodes effect magnitude (darker blue = larger improvement; orange = S-Multi is worse). Sig.\ and Mar.\ count the number of traits reaching statistical significance and marginal significance, respectively. All $p$-values are corrected for multiple comparisons using the Benjamini--Hochberg FDR procedure over 25 tests (5 single-trait conditions $\times$ 5 personality traits) within each model. $^{**}p<0.01$, $^{*}p<0.05$, $^{\dagger}p<0.1$ (corrected).}
\label{tab:paired_bfi}
\begin{tabular}{llccccccc}
\toprule
\textbf{Model} & \textbf{Condition} & \textbf{O} & \textbf{C} & \textbf{E} & \textbf{A} & \textbf{N} & \textbf{Sig.} & \textbf{Mar.} \\
\midrule
\multirow{5}{*}{Qwen3.5-Plus} & S-O & \cellcolor{pos2} 0.140 & \cellcolor{pos3} 0.188$^{\dagger}$ & \cellcolor{pos2} 0.112 & \cellcolor{pos6} \textbf{0.326}$^{*}$ & \cellcolor{pos4} 0.206$^{\dagger}$ & 1 & 2 \\
 & S-C & \cellcolor{pos4} 0.243$^{*}$ & \cellcolor{pos4} 0.233$^{\dagger}$ & \cellcolor{pos3} 0.196$^{\dagger}$ & \cellcolor{pos6} \textbf{0.310}$^{*}$ & \cellcolor{pos5} 0.258$^{*}$ & 3 & 2 \\
 & S-E & \cellcolor{pos5} 0.267$^{*}$ & 0.044 & \cellcolor{pos2} 0.108 & \cellcolor{pos6} \textbf{0.363}$^{*}$ & \cellcolor{pos3} 0.165 & 2 & 0 \\
 & S-A & \cellcolor{pos3} 0.174$^{\dagger}$ & \cellcolor{pos2} 0.137 & \cellcolor{pos2} 0.120 & \cellcolor{pos4} 0.248$^{*}$ & \cellcolor{pos5} \textbf{0.257}$^{\dagger}$ & 1 & 2 \\
 & S-N & \cellcolor{pos3} 0.167$^{\dagger}$ & \cellcolor{pos3} 0.162 & \cellcolor{pos1} 0.088 & \cellcolor{pos2} 0.147 & \cellcolor{pos6} \textbf{0.306}$^{*}$ & 1 & 1 \\
\midrule
\multirow{5}{*}{Kimi-K2.5} & S-O & \cellcolor{pos6} \textbf{0.326}$^{**}$ & \cellcolor{pos2} 0.135 & \cellcolor{pos2} 0.127 & \cellcolor{pos2} 0.118 & \cellcolor{neg} $-$0.030 & 1 & 0 \\
 & S-C & \cellcolor{pos6} \textbf{0.320}$^{**}$ & \cellcolor{pos3} 0.156 & \cellcolor{pos4} 0.243$^{*}$ & \cellcolor{pos4} 0.224$^{\dagger}$ & \cellcolor{neg} $-$0.074 & 2 & 1 \\
 & S-E & \cellcolor{pos5} \textbf{0.298}$^{**}$ & \cellcolor{pos3} 0.173 & \cellcolor{pos4} 0.232$^{*}$ & \cellcolor{pos4} 0.210 & \cellcolor{neg} $-$0.058 & 2 & 0 \\
 & S-A & \cellcolor{pos4} \textbf{0.207}$^{*}$ & 0.044 & \cellcolor{pos1} 0.072 & \cellcolor{pos2} 0.138 & 0.023 & 1 & 0 \\
 & S-N & \cellcolor{pos4} \textbf{0.229}$^{**}$ & \cellcolor{pos1} 0.077 & \cellcolor{pos1} 0.098 & \cellcolor{pos1} 0.080 & \cellcolor{pos1} 0.092 & 1 & 0 \\
\midrule
\multirow{5}{*}{GLM-5} & S-O & \cellcolor{pos7} \textcolor{white}{0.423}$^{**}$ & \cellcolor{pos4} 0.240$^{**}$ & \cellcolor{pos3} 0.188$^{*}$ & \cellcolor{pos7} \textcolor{white}{\textbf{0.469}}$^{**}$ & \cellcolor{pos6} 0.305$^{**}$ & 5 & 0 \\
 & S-C & \cellcolor{pos8} \textcolor{white}{0.518}$^{**}$ & \cellcolor{pos4} 0.201$^{*}$ & \cellcolor{pos7} \textcolor{white}{0.402}$^{**}$ & \cellcolor{pos8} \textcolor{white}{\textbf{0.537}}$^{**}$ & \cellcolor{pos5} 0.250$^{**}$ & 5 & 0 \\
 & S-E & \cellcolor{pos7} \textcolor{white}{0.411}$^{**}$ & \cellcolor{pos4} 0.229$^{*}$ & \cellcolor{pos4} 0.243$^{*}$ & \cellcolor{pos8} \textcolor{white}{\textbf{0.565}}$^{**}$ & \cellcolor{pos6} 0.331$^{**}$ & 5 & 0 \\
 & S-A & \cellcolor{pos6} 0.322$^{**}$ & \cellcolor{pos6} 0.338$^{**}$ & \cellcolor{pos2} 0.123 & \cellcolor{pos7} \textcolor{white}{\textbf{0.471}}$^{**}$ & \cellcolor{pos5} 0.299$^{**}$ & 4 & 0 \\
 & S-N & \cellcolor{pos5} 0.279$^{*}$ & \cellcolor{pos3} 0.196$^{\dagger}$ & \cellcolor{pos2} 0.137 & \cellcolor{pos5} 0.280$^{*}$ & \cellcolor{pos7} \textcolor{white}{\textbf{0.413}}$^{**}$ & 3 & 1 \\
\midrule
\multirow{5}{*}{DeepSeek-V3.2} & S-O & \cellcolor{pos7} \textcolor{white}{\textbf{0.474}}$^{**}$ & \cellcolor{pos6} 0.369$^{**}$ & \cellcolor{pos3} 0.195 & \cellcolor{pos5} 0.292$^{*}$ & \cellcolor{pos6} 0.354$^{**}$ & 4 & 0 \\
 & S-C & \cellcolor{pos8} \textcolor{white}{\textbf{0.571}}$^{**}$ & \cellcolor{pos3} 0.185 & \cellcolor{pos6} 0.355$^{**}$ & \cellcolor{pos7} \textcolor{white}{0.411}$^{**}$ & \cellcolor{pos5} 0.287$^{*}$ & 4 & 0 \\
 & S-E & \cellcolor{pos8} \textcolor{white}{\textbf{0.500}}$^{**}$ & \cellcolor{pos4} 0.219$^{*}$ & \cellcolor{pos5} 0.263$^{*}$ & \cellcolor{pos3} 0.182 & \cellcolor{pos4} 0.221$^{*}$ & 4 & 0 \\
 & S-A & \cellcolor{pos6} \textbf{0.393}$^{**}$ & \cellcolor{pos5} 0.258$^{*}$ & \cellcolor{pos2} 0.132 & \cellcolor{pos3} 0.169 & \cellcolor{pos6} 0.348$^{**}$ & 3 & 0 \\
 & S-N & \cellcolor{pos7} \textcolor{white}{\textbf{0.443}}$^{**}$ & \cellcolor{pos4} 0.217$^{*}$ & \cellcolor{pos3} 0.192 & \cellcolor{pos1} 0.093 & \cellcolor{pos5} 0.285$^{*}$ & 3 & 0 \\
\midrule
\multirow{5}{*}{GPT-5.4} & S-O & \cellcolor{pos3} 0.167$^{**}$ & \cellcolor{pos4} \textbf{0.205}$^{**}$ & \cellcolor{pos1} 0.096$^{\dagger}$ & \cellcolor{pos2} 0.120 & \cellcolor{pos3} 0.187$^{**}$ & 3 & 1 \\
 & S-C & \cellcolor{pos4} \textbf{0.202}$^{*}$ & \cellcolor{pos2} 0.148$^{*}$ & \cellcolor{pos3} 0.152$^{*}$ & \cellcolor{pos1} 0.062 & \cellcolor{pos3} 0.192$^{*}$ & 4 & 0 \\
 & S-E & \cellcolor{pos4} \textbf{0.202}$^{**}$ & \cellcolor{pos3} 0.179$^{**}$ & \cellcolor{pos2} 0.143$^{\dagger}$ & \cellcolor{pos2} 0.140 & \cellcolor{pos2} 0.132$^{\dagger}$ & 2 & 2 \\
 & S-A & \cellcolor{pos2} 0.138$^{*}$ & \cellcolor{pos4} 0.213$^{**}$ & \cellcolor{neg} $-$0.006 & \cellcolor{pos1} 0.061 & \cellcolor{pos5} \textbf{0.265}$^{**}$ & 3 & 0 \\
 & S-N & \cellcolor{pos2} 0.119$^{*}$ & \cellcolor{pos2} 0.135$^{\dagger}$ & \cellcolor{pos1} 0.076 & 0.012 & \cellcolor{pos2} \textbf{0.149}$^{*}$ & 2 & 1 \\
\midrule
\multirow{5}{*}{Claude Sonnet 4.6} & S-O & \cellcolor{pos2} 0.117 & \cellcolor{pos3} 0.180$^{*}$ & \cellcolor{pos4} 0.205$^{*}$ & \cellcolor{pos5} \textbf{0.258}$^{*}$ & \cellcolor{pos1} 0.073 & 3 & 0 \\
 & S-C & \cellcolor{pos3} 0.193$^{\dagger}$ & \cellcolor{pos2} 0.106 & \cellcolor{pos7} \textcolor{white}{\textbf{0.429}}$^{**}$ & \cellcolor{pos6} 0.370$^{**}$ & \cellcolor{pos2} 0.102 & 2 & 1 \\
 & S-E & \cellcolor{pos2} 0.129 & \cellcolor{pos4} 0.215$^{*}$ & \cellcolor{pos5} 0.286$^{**}$ & \cellcolor{pos6} \textbf{0.365}$^{*}$ & \cellcolor{pos3} 0.164$^{*}$ & 4 & 0 \\
 & S-A & \cellcolor{neg} $-$0.024 & \cellcolor{pos2} 0.119 & \cellcolor{pos2} 0.119 & \cellcolor{pos5} \textbf{0.284}$^{*}$ & \cellcolor{pos4} 0.237$^{*}$ & 2 & 0 \\
 & S-N & 0.033 & \cellcolor{pos4} 0.220$^{\dagger}$ & \cellcolor{pos2} 0.144 & 0.012 & \cellcolor{pos6} \textbf{0.320}$^{**}$ & 1 & 1 \\
\bottomrule
\end{tabular}
\end{table*}

\subsection{Model Performance using Expert Ratings as Ground Truth}
\label{app:expert_mae}

To test whether the advantage of multi-context assessment holds under an independent ground truth, we repeated the MAE analysis using ratings from four personality psychology experts as the reference (Section~4.4). Table~\ref{tab:sep_expert_mae} reports results separately for each annotator to preserve transparency.
 
S-Multi achieves the lowest MAE for every model under every annotator, without exception. Absolute MAE values are higher than those under BFI-44 (e.g., GPT-5.4 S-Multi ranges from 0.716 to 0.836 across annotators, compared with 0.607 under BFI-44). This is expected, as expert ratings and self-report measure partially overlapping but distinct aspects of personality; the gap in absolute MAE does not indicate that one reference is more valid than the other. The key finding is that the relative ordering---S-Multi outperforming all single-context conditions---is consistent across both ground truth sources.

\begin{table*}[t]
\centering
\caption{MAE of each LLM under Single-trait conditions evaluated against four human expert annotations. S-O through S-N denote Single-trait conditions and Avg denotes the mean MAE across the five Single-trait conditions. \textbf{Bold} indicates the best (lowest) value per row; \underline{underline} indicates the second best.}
\label{tab:sep_expert_mae}
\begin{tabularx}{\textwidth}{l|*{5}{>{\centering\arraybackslash}X}>{\centering\arraybackslash}p{1.1cm}|*{5}{>{\centering\arraybackslash}X}>{\centering\arraybackslash}p{1.1cm}}
\toprule
& \multicolumn{6}{c|}{\textbf{Annotator 1}} & \multicolumn{6}{c}{\textbf{Annotator 2}} \\
\midrule
\textbf{Model} & \textbf{S-O} & \textbf{S-C} & \textbf{S-E} & \textbf{S-A} & \textbf{S-N} & \textbf{S-Multi} & \textbf{S-O} & \textbf{S-C} & \textbf{S-E} & \textbf{S-A} & \textbf{S-N} & \textbf{S-Multi} \\
\hline
Qwen3.5-Plus & 0.977 & \underline{0.975} & 1.020 & 0.998 & 0.985 & \textbf{0.866} & \underline{1.021} & 1.026 & 1.055 & 1.031 & 1.027 & \textbf{0.915} \\
Kimi-K2.5 & \underline{1.092} & 1.129 & 1.118 & 1.100 & 1.127 & \textbf{0.982} & \underline{1.146} & 1.191 & 1.196 & 1.150 & 1.193 & \textbf{1.054} \\
GLM-5 & 1.094 & 1.120 & 1.188 & \underline{1.080} & 1.105 & \textbf{0.960} & 1.149 & 1.174 & 1.240 & \underline{1.131} & 1.157 & \textbf{1.008} \\
DeepSeek-V3.2 & 1.312 & 1.361 & 1.270 & \underline{1.216} & 1.263 & \textbf{1.128} & 1.390 & 1.430 & 1.353 & \underline{1.293} & 1.332 & \textbf{1.203} \\
GPT-5.4 & 0.874 & 0.851 & 0.910 & \underline{0.829} & 0.868 & \textbf{0.796} & 0.913 & 0.890 & 0.944 & \underline{0.877} & 0.901 & \textbf{0.836} \\
Claude Sonnet 4.6 & 1.074 & 1.111 & 1.118 & 1.069 & \underline{1.045} & \textbf{0.974} & 1.130 & 1.160 & 1.161 & 1.126 & \underline{1.101} & \textbf{1.024} \\
\midrule
& \multicolumn{6}{c|}{\textbf{Annotator 3}} & \multicolumn{6}{c}{\textbf{Annotator 4}} \\
\hline
Qwen3.5-Plus & 0.906 & 0.912 & 0.958 & 0.921 & \underline{0.897} & \textbf{0.776} & 0.974 & 0.968 & 1.018 & 0.977 & \underline{0.964} & \textbf{0.848} \\
Kimi-K2.5 & \underline{0.997} & 1.026 & 1.026 & 0.998 & 1.030 & \textbf{0.872} & \underline{1.070} & 1.110 & 1.102 & 1.083 & 1.106 & \textbf{0.958} \\
GLM-5 & \underline{1.004} & 1.041 & 1.113 & 1.042 & 1.025 & \textbf{0.866} & 1.081 & 1.105 & 1.185 & \underline{1.078} & 1.090 & \textbf{0.942} \\
DeepSeek-V3.2 & 1.239 & 1.274 & 1.189 & \underline{1.152} & 1.155 & \textbf{1.027} & 1.301 & 1.350 & 1.264 & \underline{1.218} & 1.238 & \textbf{1.107} \\
GPT-5.4 & 0.809 & 0.785 & 0.829 & \underline{0.772} & 0.781 & \textbf{0.716} & 0.864 & 0.843 & 0.900 & \underline{0.825} & 0.848 & \textbf{0.782} \\
Claude Sonnet 4.6 & 0.975 & 1.010 & 1.016 & 0.962 & \underline{0.925} & \textbf{0.843} & 1.052 & 1.081 & 1.101 & 1.041 & \underline{1.014} & \textbf{0.942} \\
\bottomrule
\end{tabularx}
\end{table*}

\subsection{Paired T-test results using Expert Ratings as Ground Truth}
\label{app:paired_expert}
Table~\ref{tab:paired_expert} reports paired $t$-test results using mean expert ratings as ground truth, paralleling the BFI-44-based analysis in Appendix~\ref{app:paired_bfi}.
 
The overall pattern is consistent with the BFI-44 results: the large majority of paired differences are positive, and none of the negative cells reach statistical significance after FDR correction. The number of corrected-significant comparisons is somewhat lower for certain models (e.g., GPT-5.4), which is expected given the higher variability in expert ratings (inter-rater ICC ranged from 0.567 to 0.844 across dimensions). The convergence of conclusions across two methodologically independent ground truth sources strengthens the validity of Finding~1.

\begin{table*}[t]
\centering
\definecolor{neg}{HTML}{FDE0CB}
\definecolor{pos1}{HTML}{EFF3FF}
\definecolor{pos2}{HTML}{DEEBF7}
\definecolor{pos3}{HTML}{C6DBEF}
\definecolor{pos4}{HTML}{9ECAE1}
\definecolor{pos5}{HTML}{6BAED6}
\definecolor{pos6}{HTML}{4292C6}
\definecolor{pos7}{HTML}{2171B5}
\definecolor{pos8}{HTML}{084594}
\caption{Paired $t$-test results comparing M-Concat against each Single-trait condition, using human expert evaluation as ground truth. Each cell reports the mean paired difference in MAE (Single-trait $-$ M-Concat) across participants for a given trait, where positive values indicate that M-Concat achieves lower (better) MAE. Bold denotes the largest improvement per row. Cell shading encodes effect magnitude (darker blue = larger improvement; orange = M-Concat is worse). Sig.\ and Mar.\ count the number of traits reaching statistical significance and marginal significance, respectively. All $p$-values are corrected for multiple comparisons using the Benjamini--Hochberg FDR procedure over 25 tests (5 single-trait conditions $\times$ 5 personality traits) within each model. $^{**}p<0.01$, $^{*}p<0.05$, $^{\dagger}p<0.1$ (corrected).}
\label{tab:paired_expert}
\begin{tabular}{llccccccc}
\toprule
\textbf{Model} & \textbf{Cond.} & \textbf{O} & \textbf{C} & \textbf{E} & \textbf{A} & \textbf{N} & \textbf{Sig.} & \textbf{Mar.} \\
\midrule
\multirow{5}{*}{Qwen3.5-Plus} & S-O & \cellcolor{pos2} 0.124$^{**}$ & \cellcolor{pos4} \textbf{0.235}$^{**}$ & \cellcolor{pos1} 0.097 & \cellcolor{pos4} 0.218$^{*}$ & \cellcolor{neg} $-$0.027 & 3 & 0 \\
 & S-C & \cellcolor{pos2} 0.116$^{*}$ & \cellcolor{pos3} 0.189$^{*}$ & 0.027 & \cellcolor{pos4} \textbf{0.216}$^{*}$ & \cellcolor{pos2} 0.129$^{*}$ & 4 & 0 \\
 & S-E & \cellcolor{pos3} 0.190$^{**}$ & \cellcolor{pos2} 0.119$^{\dagger}$ & \cellcolor{pos3} 0.171$^{**}$ & \cellcolor{pos6} \textbf{0.335}$^{**}$ & \cellcolor{pos1} 0.094 & 3 & 1 \\
 & S-A & \cellcolor{pos3} 0.166$^{**}$ & \cellcolor{pos4} \textbf{0.229}$^{**}$ & \cellcolor{pos1} 0.096$^{\dagger}$ & \cellcolor{pos3} 0.186$^{*}$ & 0.049 & 3 & 1 \\
 & S-N & 0.016 & \cellcolor{pos5} \textbf{0.261}$^{**}$ & \cellcolor{neg} $-$0.023 & \cellcolor{pos3} 0.189$^{*}$ & \cellcolor{pos3} 0.162$^{**}$ & 3 & 0 \\
\midrule
\multirow{5}{*}{Kimi-K2.5} & S-O & \cellcolor{pos2} 0.112$^{*}$ & \cellcolor{pos5} \textbf{0.267}$^{**}$ & \cellcolor{pos1} 0.059 & \cellcolor{pos4} 0.216$^{*}$ & \cellcolor{neg} $-$0.027 & 3 & 0 \\
 & S-C & \cellcolor{pos1} 0.054 & \cellcolor{pos4} \textbf{0.231}$^{*}$ & \cellcolor{pos1} 0.094 & \cellcolor{pos4} 0.225$^{*}$ & \cellcolor{pos1} 0.098$^{\dagger}$ & 2 & 1 \\
 & S-E & 0.049 & \cellcolor{pos3} 0.172$^{\dagger}$ & \cellcolor{pos3} 0.176$^{**}$ & \cellcolor{pos5} \textbf{0.275}$^{*}$ & \cellcolor{pos2} 0.100 & 2 & 1 \\
 & S-A & \cellcolor{pos1} 0.078 & \cellcolor{pos5} 0.260$^{**}$ & 0.006 & \cellcolor{pos5} \textbf{0.261}$^{**}$ & 0.027 & 2 & 0 \\
 & S-N & \cellcolor{neg} $-$0.026 & \cellcolor{pos6} \textbf{0.310}$^{**}$ & \cellcolor{pos1} 0.056 & \cellcolor{pos4} 0.210$^{*}$ & \cellcolor{pos4} 0.239$^{**}$ & 3 & 0 \\
\midrule
\multirow{5}{*}{GLM-5} & S-O & \cellcolor{pos3} 0.193$^{**}$ & \cellcolor{pos4} \textbf{0.247}$^{**}$ & \cellcolor{pos3} 0.193$^{**}$ & \cellcolor{pos3} 0.196$^{*}$ & \cellcolor{neg} $-$0.139 & 4 & 0 \\
 & S-C & \cellcolor{pos1} 0.096 & \cellcolor{pos4} 0.202$^{*}$ & \cellcolor{pos3} 0.159$^{*}$ & \cellcolor{pos5} \textbf{0.282}$^{**}$ & \cellcolor{pos2} 0.136$^{*}$ & 4 & 0 \\
 & S-E & \cellcolor{pos5} 0.298$^{**}$ & \cellcolor{pos4} 0.230$^{*}$ & \cellcolor{pos4} 0.220$^{**}$ & \cellcolor{pos6} \textbf{0.341}$^{**}$ & \cellcolor{pos2} 0.147$^{**}$ & 5 & 0 \\
 & S-A & \cellcolor{pos3} 0.189$^{*}$ & \cellcolor{pos4} 0.214$^{*}$ & \cellcolor{pos2} 0.137$^{\dagger}$ & \cellcolor{pos5} \textbf{0.299}$^{**}$ & \cellcolor{neg} $-$0.009 & 3 & 1 \\
 & S-N & \cellcolor{pos1} 0.075 & \cellcolor{pos5} \textbf{0.272}$^{**}$ & \cellcolor{pos1} 0.097 & \cellcolor{pos3} 0.165 & \cellcolor{pos3} 0.187$^{*}$ & 2 & 0 \\
\midrule
\multirow{5}{*}{DeepSeek-V3.2} & S-O & \cellcolor{pos2} 0.107 & \cellcolor{pos6} 0.387$^{**}$ & \cellcolor{pos1} 0.089 & \cellcolor{pos7} \textcolor{white}{\textbf{0.424}}$^{**}$ & \cellcolor{pos1} 0.052 & 2 & 0 \\
 & S-C & \cellcolor{pos3} 0.169$^{*}$ & \cellcolor{pos6} 0.323$^{**}$ & 0.046 & \cellcolor{pos7} \textcolor{white}{\textbf{0.460}}$^{**}$ & \cellcolor{pos3} 0.174$^{**}$ & 4 & 0 \\
 & S-E & 0.027 & \cellcolor{pos5} 0.265$^{*}$ & \cellcolor{pos3} 0.152$^{*}$ & \cellcolor{pos6} \textbf{0.304}$^{*}$ & \cellcolor{pos1} 0.066 & 3 & 0 \\
 & S-A & 0.046 & \cellcolor{pos6} \textbf{0.316}$^{**}$ & 0.004 & \cellcolor{pos5} 0.272$^{*}$ & \cellcolor{neg} $-$0.009 & 2 & 0 \\
 & S-N & 0.028 & \cellcolor{pos4} 0.235$^{*}$ & 0.016 & \cellcolor{pos5} \textbf{0.254}$^{*}$ & \cellcolor{pos2} 0.110 & 2 & 0 \\
\midrule
\multirow{5}{*}{GPT-5.4} & S-O & \cellcolor{pos1} 0.096$^{*}$ & \cellcolor{pos1} 0.081$^{*}$ & \cellcolor{pos1} 0.053 & \cellcolor{pos3} \textbf{0.194}$^{*}$ & 0.039 & 3 & 0 \\
 & S-C & \cellcolor{neg} $-$0.013 & 0.040 & \cellcolor{pos1} 0.068 & \cellcolor{pos2} 0.120 & \cellcolor{pos2} \textbf{0.132}$^{*}$ & 1 & 0 \\
 & S-E & \cellcolor{pos1} 0.069 & \cellcolor{pos2} 0.108$^{*}$ & \cellcolor{pos2} 0.106$^{*}$ & \cellcolor{pos4} \textbf{0.200}$^{\dagger}$ & \cellcolor{pos1} 0.084 & 2 & 1 \\
 & S-A & \cellcolor{pos1} 0.087$^{\dagger}$ & 0.035 & \cellcolor{neg} $-$0.002 & \cellcolor{pos2} \textbf{0.139} & 0.022 & 0 & 1 \\
 & S-N & \cellcolor{neg} $-$0.052 & 0.043 & 0.029 & \cellcolor{pos4} \textbf{0.238}$^{*}$ & \cellcolor{pos1} 0.069 & 1 & 0 \\
\midrule
\multirow{5}{*}{Claude Sonnet 4.6} & S-O & \cellcolor{pos3} 0.164$^{**}$ & \cellcolor{pos4} \textbf{0.242}$^{**}$ & \cellcolor{pos1} 0.083 & \cellcolor{pos4} 0.214$^{*}$ & \cellcolor{neg} $-$0.040 & 3 & 0 \\
 & S-C & \cellcolor{pos2} 0.128$^{\dagger}$ & \cellcolor{pos3} 0.154$^{*}$ & \cellcolor{pos5} \textbf{0.256}$^{**}$ & \cellcolor{pos4} 0.245$^{*}$ & \cellcolor{pos1} 0.055 & 3 & 1 \\
 & S-E & \cellcolor{pos2} 0.125$^{*}$ & \cellcolor{pos3} 0.191$^{*}$ & \cellcolor{pos2} 0.140$^{*}$ & \cellcolor{pos5} \textbf{0.296}$^{*}$ & \cellcolor{pos2} 0.111$^{*}$ & 5 & 0 \\
 & S-A & \cellcolor{pos3} 0.162$^{*}$ & \cellcolor{pos2} 0.121$^{\dagger}$ & 0.034 & \cellcolor{pos5} \textbf{0.260}$^{*}$ & 0.018 & 2 & 1 \\
 & S-N & \cellcolor{neg} $-$0.024 & \cellcolor{pos3} 0.153$^{\dagger}$ & \cellcolor{neg} $-$0.023 & \cellcolor{pos2} 0.140 & \cellcolor{pos3} \textbf{0.166}$^{**}$ & 1 & 1 \\
\bottomrule
\end{tabular}
\end{table*}

\end{document}